\documentclass{aa}  

\usepackage[colorlinks=true,allcolors=blue]{hyperref}
\usepackage{threeparttable}

\usepackage{graphicx}
\usepackage{txfonts}
\usepackage{xspace}
\newcommand{\kms}{km~s$^{-1}$\xspace}
\newcommand{\ergs}{erg~s$^{-1}$\xspace}
\newcommand{\Msunyr}{M$_{\odot}$~yr$^{-1}$\xspace}

\newcommand{\hb}{H$\beta$\xspace}
\newcommand{\ha}{H$\alpha$\xspace}
\newcommand{\heii}{He\,{\sc{ii}}\xspace}

\newcommand{\oiii}{[O\,{\sc{iii}}]\xspace}

\newcommand{\siuii}{Si\,{\sc{ii}}\xspace}
\newcommand{\siuiii}{Si\,{\sc{iii}}\xspace}
\newcommand{\siiv}{Si\,{\sc{iv}}\xspace}
\newcommand{\sii}{[S\,{\sc{ii}}]\xspace}

\newcommand{\civ}{C\,{\sc{iv}}\xspace}
\newcommand{\ciii}{C\,{\sc{iii}}]\xspace}
\newcommand{\cii}{C\,{\sc{ii}}\xspace}
\newcommand{\nv}{N\,{\sc{v}}\xspace}
\newcommand{\niv}{N\,{\sc{iv}]}\xspace}
\newcommand{\nii}{[N\,{\sc{ii}]}\xspace}
\newcommand{\niii}{N\,{\sc{iii}]}\xspace}

\newcommand{\MOKA}{\ensuremath{\mathrm{MOKA^{3D}}}\xspace}

\begin{document}

   \title{GA-NIFS: A galaxy-wide outflow in a Compton-thick mini-BAL quasar at $z=3.5$ probed in emission and absorption}

   \author{Michele Perna
          \inst{\ref{iCAB}}\thanks{e-mail: mperna@cab.inta-csic.es}
          \and
          Santiago Arribas\inst{\ref{iCAB}}
          \and 
          Xihan Ji\inst{\ref{iKav},\ref{iCav}}
          \and 
          Cosimo Marconcini\inst{ \ref{iUNIFI}, \ref{iOAA}}
          \and 
          Isabella Lamperti\inst{\ref{iUNIFI}, \ref{iOAA}, \ref{iCAB}}
          \and
          Elena Bertola\inst{\ref{iOAA}}
          \and
          Chiara~Circosta\inst{\ref{iESAes}}
          \and
          Francesco D'Eugenio\inst{\ref{iKav},\ref{iCav}}
          \and
          Hannah \"{U}bler\inst{\ref{iMPE}}
          \and
          Torsten~Böker\inst{\ref{iESAusa}}
          \and
          Roberto~Maiolino\inst{\ref{iKav}, \ref{iCav},\ref{iUCL}}
          \and
          Andrew~J.~Bunker\inst{\ref{iOxf}}
          \and
          Stefano~Carniani\inst{\ref{iNorm}}
          \and
          St\'ephane Charlot\inst{\ref{iSor}}
          \and
          Chris~J.~Willott\inst{\ref{iNRC}}
          \and
          Giovanni Cresci\inst{\ref{iOAA}}
          \and
          Eleonora~Parlanti\inst{\ref{iNorm}}
          \and
          Bruno Rodr\'iguez~Del~Pino\inst{\ref{iCAB}}
          \and 
          Jan Scholtz\inst{\ref{iKav},\ref{iCav}}
          \and
          Giacomo Venturi\inst{\ref{iNorm}}
          }

   \titlerunning{GA-NIFS: A galaxy-wide outflow in a Compton thick quasar at $z=3.5$}
   \authorrunning{M. Perna et al.}
   \institute{
            Centro de Astrobiolog\'ia (CAB), CSIC--INTA, Cra. de Ajalvir Km.~4, 28850 -- Torrej\'on de Ardoz, Madrid, Spain\label{iCAB}
    \and
            Kavli Institute for Cosmology, University of Cambridge, Madingley Road, Cambridge, CB3 0HA, UK\label{iKav}
    \and
            Cavendish Laboratory - Astrophysics Group, University of Cambridge, 19 JJ Thomson Avenue, Cambridge, CB3 0HE, UK\label{iCav}
    \and
            Università di Firenze, Dipartimento di Fisica e Astronomia, via G. Sansone 1, 50019 Sesto F.no, Firenze, Italy\label{iUNIFI}
    \and 
            INAF - Osservatorio Astrofisico di Arcetri, Largo E. Fermi 5, I-50125 Firenze, Italy\label{iOAA}
    \and 
            European Space Agency, ESAC, Villanueva de la Ca\~{n}ada, E-28692 Madrid, Spain\label{iESAes}
    \and
            $^{}$Max-Planck-Institut f\"ur extraterrestrische Physik (MPE), Gie{\ss}enbachstra{\ss}e 1, 85748 Garching, Germany\label{iMPE}
    \and    
            European Space Agency, c/o STScI, 3700 San Martin Drive, Baltimore, MD 21218, USA\label{iESAusa}   
    \and
            Department of Physics, University of Oxford, Denys Wilkinson Building, Keble Road, Oxford OX1 3RH, UK\label{iOxf}
    \and
            Department of Physics and Astronomy, University College London, Gower Street, London WC1E 6BT, UK\label{iUCL}  
    \and
            Scuola Normale Superiore, Piazza dei Cavalieri 7, I-56126 Pisa, Italy\label{iNorm}
    \and
            Sorbonne Universit\'e, CNRS, UMR 7095, Institut d’Astrophysique de Paris, 98 bis bd Arago, 75014 Paris, France\label{iSor} 
    \and 
            National Research Council of Canada, Herzberg Astronomy \& Astrophysics Research Centre, 5071 West Saanich Road, Victoria, BC V9E 2E7, Canada\label{iNRC}
        }


 
  \abstract
   {Studying the distribution and properties of ionised gas in outflows driven by active galactic nuclei (AGN) is crucial for understanding the feedback mechanisms at play in extragalactic environments. These outflows provide key insights into the regulation of star formation and the growth of supermassive black holes.
   }
   {In this study, we explore the connection between ionised outflows traced by rest-frame ultra-violet (UV) absorption and optical emission lines in GS133, a Compton thick AGN at $z = 3.47$.  
   We combine observations from the James Webb Space Telescope (JWST) NIRSpec Integral Field Spectrograph (IFS) with archival Very Large Telescope (VLT) VIMOS long-slit spectroscopic data, as part of the ``Galaxy Assembly with NIRSpec IFS'' (GA-NIFS) project.}
   {We perform a multi-component kinematic decomposition of the UV and optical line profiles to derive the physical properties of the absorbing and emitting gas in  GS133.}
   {Our kinematic decomposition reveals two distinct components in the optical emission lines. The first component likely traces a rotating disk with a dynamical mass of $2\times 10^{10}$~M$_\odot$. The second component corresponds to a galaxy-wide, bi-conical outflow, with a velocity of $\sim \pm 1000$~\kms and an extension of $\sim 3$~kpc.  
   The UV absorption lines show two outflow components, with bulk velocities $v_{\rm out} \sim -900$~\kms and $\sim -1900$~\kms, respectively. This characterizes GS133 as a mini-broad absorption line (mini-BAL) system. Balmer absorption lines with similar velocities are tentatively detected in the NIRSpec spectrum. Both photoionisation models and outflow energetics suggest that the ejected absorbing gas is located at 1--10~kpc from the AGN. We use 3D gas kinematic modelling to infer the orientation of the \oiii bi-conical outflow, and find that a portion of the emitting gas resides along our line of sight, suggesting that \oiii and absorbing gas clouds are partially mixed in the outflow. The derived  
   mass-loading factor (i.e. the mass outflow rate divided by the star formation rate) of 1--10, and the kinetic coupling efficiency (i.e. the kinetic power divided by L$_{\rm AGN}$) of 0.1--1\% suggest that the outflow in GS133 provides significant feedback on galactic scales.
  }
   {}

   \keywords{galaxies: high-redshift -- galaxies: active -- (galaxies:) quasars: supermassive black holes -- (galaxies:) quasars: absorption lines
               }

   \maketitle
%

\section{Introduction}

Accretion-disk outflows from active galactic nuclei (AGN) are thought to provide significant kinetic-energy feedback to their host galaxies, potentially stopping star formation and suppressing accretion onto the central black hole (e.g. \citealt{DiMatteo2005, Hopkins2008, Perna2018b, Giustini2019, Bertola2024}). The effectiveness of AGN feedback depends on the efficiency with which these outflows couple to the interstellar medium (ISM) in the host galaxies, an issue that remains complex and not fully understood (e.g. \citealt{Harrison2017}). 

AGN outflows are often studied through blueshifted broad absorption lines (BALs) tracing ionised gas in rest-frame ultraviolet (UV) spectra. BALs are typically defined by having a full width at half maximum (FWHM) $>3000$~\kms and velocity shifts greater than 5000~\kms in the AGN host galaxy rest frame (e.g. \citealt{Weymann1981}). Mini-BALs, narrower than BALs but still broad enough to be identified as AGN outflows, have velocity shifts between 500 and 2000~\kms (e.g. \citealt{Hamann2004,Moravec2017, Hamann2019, Maiolino2024Natur}). Even slower outflows are observed in the form of narrow absorption lines (NALs), with FWHM~$=100-500$~\kms and velocity offsets of a few 100~\kms (e.g. \citealt{Vestergaard2003, Hamann2004}). 
There is no consensus about the spatial extent of UV absorption line outflows, as they trace gas along our LOS, making direct measurements of their extent impractical. Indirect methods, such as photoionization analysis, indicate that
UV outflows can reach distances ranging from tens to even thousands of parsecs (\citealt{Arav2013, Xu2020}).  
Combined, the NAL and mini-BAL outflows are more common than the well-studied BALs in AGN spectra. Specifically, $\sim 50$\% of AGN contain either NAL or mini-BAL outflows, whereas only $\sim~20$\% contain a \civ BAL outflow (e.g. \citealt{Giustini2019} and references therein).

Another approach to studying AGN outflows involves observations of rest-frame optical emission lines, which may reveal blueshifted broad wings in  ionized gas tracers such as \nii $\lambda 6583$ and, more frequently, the \oiii $\lambda 5007$ forbidden lines (e.g. \citealt{Arribas2014, Genzel2014, Perna2015a, Perna2019, Perrotta2019,ForsterSchreiber2019, Kakkad2020, Tozzi2024}). 
Since forbidden features can originate only from low electron density regions (i.e. $n_e < 10^6$~cm$^{-3}$), their broadening cannot be explained by any contamination from high density AGN broad line regions (BLR).  Therefore, forbidden lines are an excellent tracer of ionised outflows, on scales extending up to several kiloparsecs from the AGN (e.g. \citealt{Harrison2012, Liu2013, Cresci2023, Ulivi2024a}). A large fraction of AGN presents signatures of outflows in their \oiii phase, from $\sim 20$\% to $\sim 70$\%, depending on the selection criteria and outflow velocity thresholds used (e.g. \citealt{VeronCetty2001,Mullaney2013, Perna2017a, Musiimenta2023}).

More generally, ejections of material from the inner regions up to the host galaxy scale can involve different gas phases, and are revealed as parsec-scale BALs (e.g. \citealt{Bruni2019, Vietri2022}) and ultrafast outflows (up to $\sim 30$\% of the speed of light)  detected in X-rays (e.g. \citealt{Pounds2003, Chartas2021, Matzeu2023}), to kiloparsec-scale outflows observed in molecular, neutral, and
ionised gas (e.g. \citealt{Cicone2014, Lamperti2022, Mehdipour2023, Parlanti2024b, Ulivi2024b}). Therefore, a comprehensive characterization of the outflow phenomena requires the use of a range of major facilities that work at different wavelengths and angular resolutions. 

The need of multiple observations with different facilities 
have affected the direct comparison of outflows identified with different tracers, like the kinematically disturbed UV absorption- and optical emission-line components. 
It is not yet determined whether these phenomena are entirely distinct and are caused during different phases in the AGN fuelling and outflow life cycle, or whether they represent the same outflow episode observed with different methods, possibly depending on the inclination of the systems relative to the line of sight (LOS). In fact, individual absorption lines only reveal the outflow components along the LOS to the AGN; in contrast, optical emission lines can trace outflow emission in any direction.
The literature includes only a limited number of studies that compare absorption and emission lines across AGN samples (e.g. \citealt{Schulze2017, Xu2020, Ahmed2024,  Temple2024, Kehoe2024}). 

Analyses of UV absorption and optical forbidden emission lines in individual AGN sources are also rare: 
there is only one integral field spectroscopy (IFS) study of two bright AGN at $z \lesssim 0.1$ (\citealt{Liu2015}), and long-slit works for three individual quasars at $z\approx 2$ (\citealt{Tian2019,Dai2024,Stepney2024}). 
The use of IFS observations is particularly important as they allow mapping of the ejected \oiii gas in two spatial and one velocity dimension. This capability enables direct measurement of the spatial extent of the outflows, which can only be indirectly inferred from UV absorption line analyses. In turn, this information is essential to quantify the impact these outflows have on the host galaxy.

In this paper, we present the James Webb Space Telescope (JWST) Near Infrared Spectrograph (NIRSpec) IFS and Very Large Telescope (VLT) VIMOS observations of GS133, a highly obscured AGN at $z = 3.5$. The NIRSpec IFS data covers the rest-frame optical ionised gas, while the VIMOS long-slit spectrum covers the rest-frame UV features. Therefore, the combination of these datasets allows us to characterise both UV and optical outflows. 
GS133 is located at 03:32:04.938, --27:44:31.73 in the GOODS-South field (\citealt{Giavalisco2004}), and is classified as an X-ray AGN in the 7 Ms exposure of the Chandra Deep Field-South (target \#133 in \citealt{Luo2017}). Both X-ray spectroscopy fitting analysis (\citealt{Luo2017}) and mid-infrared diagnostics (\citealt{Guo2021}) identify this source as a Compton thick (CT) AGN, having a column density N$_{\rm H} > 1.5\times 10^{24}$~cm$^{-2}$ (\citealt{Comastri2004}). It has an absorption-corrected 2--10~keV luminosity L$_{\rm X} = 9\times 10^{43}$~\ergs, an integrated luminosity from UV to infrared L$_{\rm bol} = 4.9\times 10^{45}$~\ergs, a black hole mass M$_{\rm BH} = 2\times 10^7$~M$_\odot$, and an Eddington ratio $\lambda_{\rm Edd} = {\rm L_{bol}/L_{Edd}}= 2.4$ (\citealt{Guo2020}). Note that because GS133 is a highly obscured AGN with no BLR emission, the BH mass was estimated assuming the M$_{\rm BH}$--M$_*$ relation with a scaling of 0.003 (\citealt{Guo2021}). 

The paper is outlined as follows. In Section \ref{sec:observations} we describe our JWST/NIRSpec IFS observations and data reduction, as well as the archival VLT/VIMOS data. Detailed data analysis of the optical and UV integrated spectra are reported in Sect. \ref{sec:spatiallyintegrateddata}, while the NIRSpec spatially resolved spectroscopic analysis is reported in Sect. \ref{sec:spatiallyresolveddata}.
Multi-wavelength diagnostics are discussed in Sect. \ref{sec:diagnostics}. The calculation of the dynamical mass of the GS133 host galaxy is provided in Sect. \ref{sec:mdyn}, followed by an analysis of the outflow energetics in Sect. \ref{sec:energetics}, which incorporates photoionisation and 3D gas kinematic models. The GS133 environment is examined in Sect. \ref{sec:environment}.
A summary of our findings is presented in Sect. \ref{sec:conclusions}.
Throughout, we adopt a \cite{Chabrier2003} initial mass function ($0.1-100~M_\odot$) and a flat $\Lambda$CDM cosmology with $H_0=70$~km~s$^{-1}$~Mpc$^{-1}$, $\Omega_\Lambda=0.7$, and $\Omega_m=0.3$.
In our analysis of NIRSpec data, we use vacuum wavelengths according to their calibration. However, when discussing rest-frame optical emission lines, we quote their air wavelengths. Similarly, for VIMOS data, we use the air frame  based on their wavelength calibration, but refer to rest-frame UV lines using their vacuum wavelengths, consistent with the standard pre-JWST practice.


\begin{figure*}[!htb]
\centering
\includegraphics[width=\textwidth]{{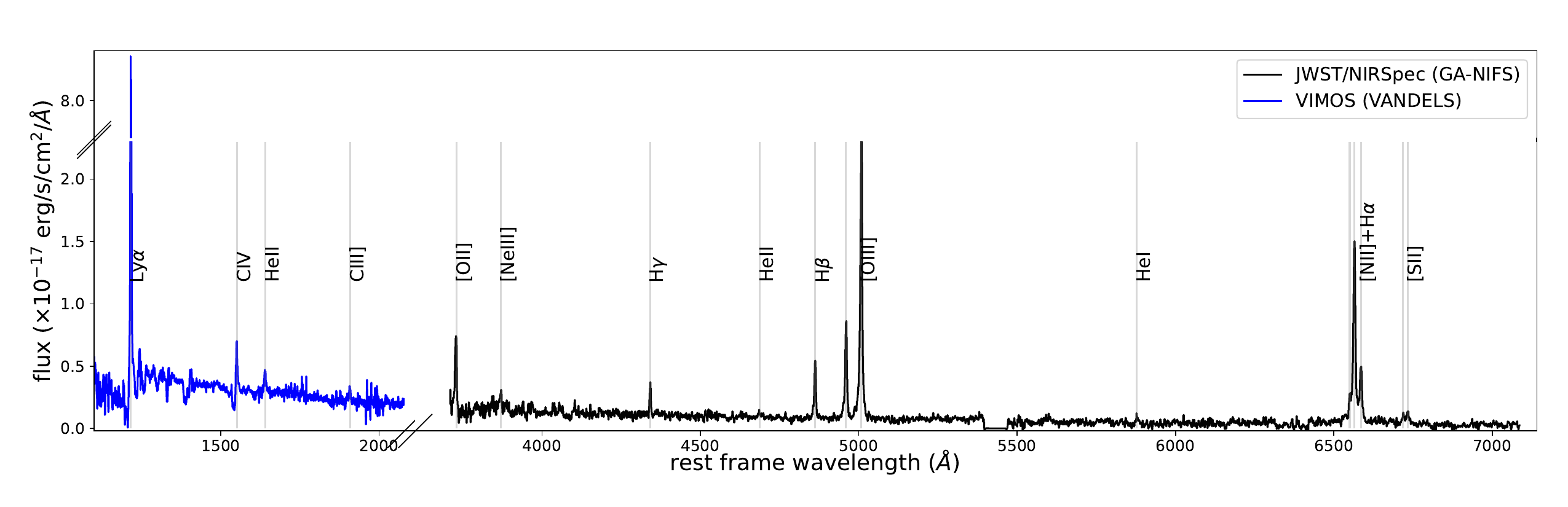}}
\caption{Integrated spectrum of GS133. The blue curve identifies the VLT/VIMOS spectrum;  the black line shows the JWST/NIRSpec spectrum, integrated over a circular aperture of $r = 0.5$\arcsec. The most prominent emission lines are marked with grey vertical lines. The gap in the middle of the NIRSpec spectrum ($\lambda_{\rm rest}\sim 5400$~\AA) is due to the separation between the two NIRSpec detectors.}\label{fig:integratedspectra}
\end{figure*}

\section{Observations and data processing}\label{sec:observations}

\subsection{NIRSpec data}
GS133  was observed on September 12$^{th}$ 2022, under program \#1220 (PI: N. L\"utzgendorf), as part of the NIRSpec IFS GTO program ``Galaxy Assembly with NIRSpec IFS'' (GA-NIFS, e.g. \citealt{RodriguezDelPino2024,  Scholtz2024cos30, Lamperti2024}). The project is based on the use of the NIRSpec’s IFS mode, which provides spatially resolved spectroscopy over a contiguous 3.1$^{\prime\prime} \times$ 3.2$^{\prime\prime}$ sky area, with a sampling of 0.1$^{\prime\prime}$/spaxel and a comparable spatial resolution (\citealt{Boker2022, Rigby2023}). 
The IFS observations were taken with the grating/filter pair G235H/F170LP. This results in a data cube with spectral resolution $R\sim2700$ over the wavelength range 1.7--3.1 $\mu$m \citep{Jakobsen2022}.
The observations were taken with the  NRSIRS2RAPID readout pattern (\citealt{Rauscher2017}) with 60 groups per integration and one integration per exposure, using a 4-point medium cycling dither pattern, resulting in a total exposure time of 3560 seconds.

We used v1.8.2 of the JWST pipeline with CRDS context 1105 to create a final cube with drizzle weighting. A patch was included to correct some important bugs that affect this specific version of the pipeline (see details in \citealt{Perna2023}). We corrected count-rate images for 1/f noise through a polynomial fit. During stage 2, we removed all data in regions of known failed open MSA shutters. We also masked pixels at the edge of the slices (two pixel wide) to conservatively exclude pixels with unreliable sflat corrections, and implemented the outlier rejection of \cite{DEugenio2023}. The combination of a dither and drizzle weighting allowed us to sub-sample the detector pixels, resulting in cube spaxels of 0.05\arcsec.

\subsection{VIMOS data}

The rest-frame UV spectrum of GS133 was observed as part of the survey ``VANDELS: a VIMOS survey of the CDFS and UDS fields'' (\citealt{Mclure2018}). This ESO public spectroscopic survey, conducted with the VIMOS spectrograph on the VLT, aimed to obtain ultradeep, medium resolution, red-optical spectra of $\sim 2000$ galaxies at $1<z<7$. 

GS133 was observed on 28 May 2017, using the MR grism and the GG475 filter, with a 1\arcsec\ slit width and a slit length of 10\arcsec\ oriented east-west on the sky. The total exposure time was 41 hours. This setup provides a wavelength coverage of 4800–10000~$\AA$ with a nominal resolution R~$= 580$, corresponding to a velocity resolution of approximately 500~\kms. The spectroscopic analysis presented in this work takes advantage of the VANDELS DR4 fully reduced spectra (\citealt{Garilli2021}), downloaded from the ESO archive\footnote{\url{https://archive.eso.org/dataset/ADP.2021-02-01T16:09:36.529}}.  

A visual inspection of the 2-dimensional spectrum does not reveal any spatially resolved structure. Therefore, the analysis presented in this work focusses on the one-dimensional (1D) VIMOS spectrum.

\section{Spatially integrated spectra}\label{sec:spatiallyintegrateddata}

Figure \ref{fig:integratedspectra} shows the integrated VIMOS and NIRSpec spectra of GS133. The VIMOS spectrum (blue curve) covers the $1070-2300~\AA$ rest-frame wavelength range, while the NIRSpec spectrum (black curve) covers the range between 3710 and 7085~$\AA$ rest-frame; it was extracted from a circular aperture centred at the position of the AGN, with a radius $r= 0.5\arcsec$, which matches the width of the VIMOS slit. In this section, we analyse these spectra, and infer the integrated properties of the outflows both in the UV range, which is spatially unresolved in VIMOS data, and in optical, making use of NIRSpec IFS observations. Instead, Sect. \ref{sec:spatiallyresolveddata} investigates the spatially resolved properties of the optical emission.

\begin{figure*}[!htb]
\centering
\includegraphics[width=\textwidth]{{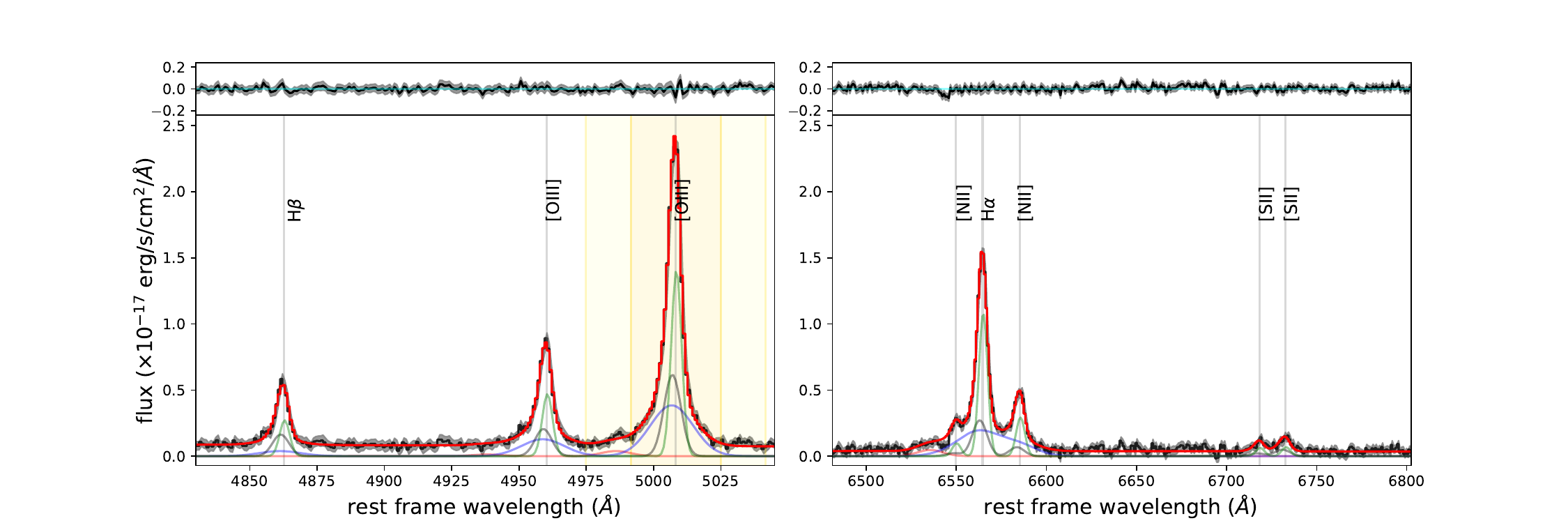}}
\caption{ JWST/NIRSpec integrated spectrum of GS133. The black curve identifies the integrated spectrum; the total, multi-component best-fit curve is in red, while all individual Gaussian components are shown with different colours. The fit residuals are reported in the top panels. The most prominent emission lines are marked with grey vertical lines. 
Light and dark yellow shaded areas in the right panel mark the \oiii emission within $\pm 2000$~\kms and $\pm 1000$~\kms, respectively, from the GS133 systemic redshift.  
}\label{fig:opticalintegratedspectrum}
\end{figure*}

\begin{table}[h]
\centering
\caption{Properties of GS133 inferred from the optical emission lines of the NIRSpec integrated spectrum, obtained for the systemic (`sys'), outflow (`out'), and total (`tot') line profiles.}%
\begin{tabular}{lc}
\hline

Measurement  & Value\\
\hline
\hline

L$_{\rm [OIII], \ sys}$ & $(1.21\pm 0.02)\times 10^{43}$~\ergs\\
L$_{\rm H\alpha, \ sys}$ & $(9.97\pm 0.13)\times 10^{42}$~\ergs\\
A$_{\rm V, \ sys}$ & $1.3_{-0.4}^{+0.2}$ mag \\
n$_{\rm e, \ sys}$ & $3300\pm 500$~cm$^{-3}$\\

$z_{\rm sys}$ & $3.4732\pm 0.0001$\\
$\sigma_{\rm sys}$ & $137\pm 5$~\kms\\

\hline

L$_{\rm [OIII], \ out}$ & $1.05_{-0.04}^{+0.01}\times 10^{43}$~\ergs\\
L$_{\rm H\alpha, \ out}$ & $(3.4\pm 0.3)\times 10^{42}$~\ergs\\
A$_{\rm V, \ out}$ & $1.3_{-0.9}^{+0.6}$\\

$\delta v_{\rm out}$ & $-150\pm 6$~\kms\\
$\sigma_{\rm out}$ & $515\pm3$~\kms \\

$W80_{\rm out}$ & $1320\pm 60$~\kms\\
$v10_{\rm out}$ & $-890\pm 40$~\kms\\
$v90_{\rm out}$ & $425\pm 10$~\kms\\

\hline
L$_{\rm [OIII], \ tot}$  & $(2.26\pm 0.01)\times 10^{43}$~\ergs \\
L$_{\rm H\alpha, \ tot}$ & $(1.40_{-0.09}^{+0.1})\times 10^{43}$~\ergs\\
A$_{\rm V, \ tot}$ & $1.1_{-0.6}^{+0.4}$~mag\\

$\Delta v_{\rm tot}$ & $-85\pm 4$~\kms\\
$\sigma_{\rm tot}$ & $370\pm3$~\kms \\

$W80_{\rm tot}$ & $780\pm 60$~\kms\\
$v10_{\rm tot}$ & $-530\pm 30$~\kms\\
$v90_{\rm tot}$ & $250\pm 20$~\kms\\

\hline

\end{tabular} 
\tablefoot{Emission line luminosities are not corrected for dust extinction. All velocity measurements refer to the \oiii line.}   
\label{tab:OPTproperties}
\end{table}

\subsection{Optical emission lines in NIRSpec cube}\label{sec:spatiallyintegratedNIRSpecdata}

We fitted the most prominent gas emission lines by using the Levenberg-Marquardt least-squares fitting code CAP-MPFIT (\citealt{Cappellari2017}). 
In particular, we modelled the \ha and \hb lines, the \oiii $\lambda\lambda$4959,5007, \nii $\lambda\lambda$6548,83, and \sii $\lambda\lambda$6716,31 doublets with a combination of Gaussian profiles, applying a simultaneous fitting procedure, so that all line features of a given kinematic component have the same velocity centroid and FWHM (e.g. \citealt{Perna2020}). We used rest-frame vacuum wavelengths. Moreover, the relative flux of the
two \nii and \oiii doublet components was fixed to 2.99 and the \sii flux ratio was required to be within the range $0.44 <
f(\lambda 6716)/ f(\lambda 6731) < 1.42$ (\citealt{Osterbrock2006}). The final number of kinematic components used to model the spectra was derived on the basis of the Bayesian information criterion (BIC; \citealt{Schwarz1978}). 

Figure \ref{fig:opticalintegratedspectrum} shows the best-fit model around the \hb-\oiii and \ha-\nii regions. All emission lines exhibit a narrow core, and prominent blue and red wings. Modelling these complex profiles required four Gaussian components. The Gaussian component associated with the line cores likely traces the systemic, unperturbed emission in the GS133 host galaxy, with a broadening (velocity dispersion $\sigma = 137\pm 5$~\kms) attributed solely to the gravitational potential of the system.  In contrast, the other components likely trace a powerful outflow. Therefore, to characterise the outflow, we derived its properties from the integrated spectrum by considering the entire reconstructed profile, combining all but the narrowest and most prominent Gaussian component. This approach is justified by the potential blending of distinct (approaching and receding) outflow components that contribute to the overall profile (see Sect. \ref{sec:spatiallyresolveddata}).

As the measurement of kinematic properties of emission line profiles can be obtained using different methods, depending on the specific needs and characteristics of the data, in Table \ref{tab:OPTproperties} we reported both non-parametric velocities (e.g. \citealt{Zakamska2014}), obtained by measuring the velocity $v$ at which a given fraction of the total best-fit line flux is collected using a cumulative function, and the velocity offset $\Delta v$ and velocity dispersion $\sigma$ defined as the moment-1 and moment-2 measurements, respectively. Table \ref{tab:OPTproperties} presents all velocity measurements inferred from the \oiii profile, for the reconstructed narrow and outflow components, and for the best-fit total emission line profile. Specifically, we reported the 10th-percentile ($v10$) and the 90th-percentile ($v90$) velocities of the fitted line profile, associated with the highest blueshifted and redshifted gas emission, respectively; $W80$, defined as $v90-v10$ and representing the line width; and the velocity shift and velocity dispersion.

We derived the GS133 redshift from the measured wavelength of the narrow Gaussian reproducing the \oiii core emission in the integrated spectrum shown in Fig. \ref{fig:opticalintegratedspectrum}: $z = 3.4732 \pm 0.0001$. This value is roughly in agreement with previous estimates $z= 3.462$  (\citealt{Cristiani2000}) and $z = 3.4739$ (\citealt{Balestra2010}) obtained from (noisier and lower spectral resolution) ground-based rest-frame UV spectra.

The \oiii perturbed gas has a line width $\sigma_{\rm out}~= 515 \pm3$~\kms, which is $\sim 4$ times broader than the systemic component, and prominent blue wings extending to $v < -1000$~\kms (light-yellow region in Fig. \ref{fig:opticalintegratedspectrum}). 
The inferred outflow velocity for the \oiii line is consistent with the general L$_{\rm AGN}-v_{\rm out}$ trends reported in the literature (\citealt{Villar2020, Musiimenta2023}). 
Specifically, it appears to be on the higher end of the distribution. Therefore, our results support the recent findings by \citet{Tozzi2024} showing that highly obscured and CT AGN tend to exhibit faster outflows than unobscured AGN at a given L$_{\rm AGN}$ (see also Bertola et al., in prep.). A similar conclusion could be reached for another highly obscured QSO in our GA-NIFS survey, ALESS073.1, presented in \citet{Parlanti2024a}. 
The GS133 \oiii outflow is investigated in more detail in Sect. \ref{sec:spatiallyresolveddata}.

\begin{figure*}[!htb]
\centering
\includegraphics[width=\textwidth]{{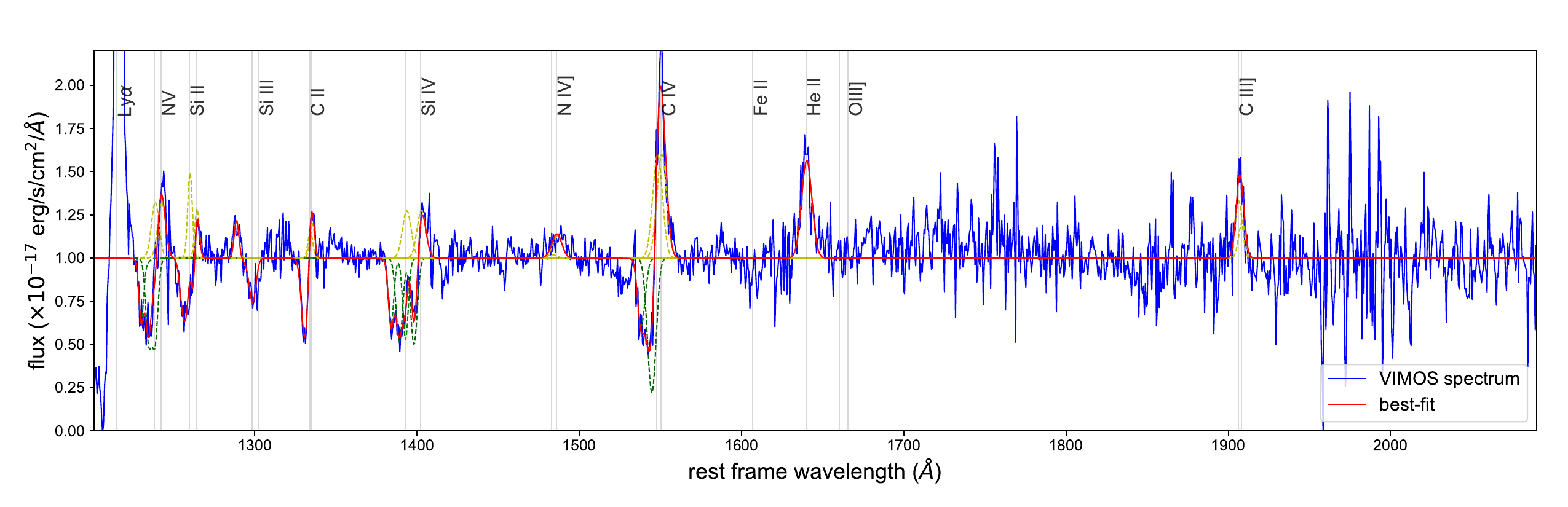}}
\caption{ VLT/VIMOS spectrum with best-fit results. The flux is normalised to the continuum. The blue curve identifies the integrated spectrum; the total, multi-component best fit curve
is in red, while all individual Gaussian (Voigt) components are shown in yellow (green). The most prominent line transitions are marked with grey vertical lines.  }\label{fig:VIMOS_EM}
\end{figure*}

\subsection{UV lines in VIMOS spectrum}

UV lines were fitted with a multi-step approach, that we summarize here. In the next section we describe our approach in full detail. First, we shifted the wavelength axis to the rest frame, using the redshift inferred from the NIRSpec data. Then, we fitted the continuum emission at $> 1220~\AA$ with a power-law, by masking all prominent line features (both in absorption and emission). Next, we normalized the continuum to 1 by dividing the original spectrum by the best-fit power-law profile. We subsequently fitted the emission lines with multi-component Gaussian profiles, and the absorption contribution with Voigt profiles. Finally, we obtained a synthetic spectrum without emission line contribution, by subtracting the best-fit emission line Gaussian profiles. This synthetic spectrum, only containing the contribution from absorption features, was finally ingested to VoigtFit (\citealt{Krogager2018}), to obtain the main physical properties of the various atomic species, namely the column density and kinematics. 

Because the VANDELS spectra are calibrated in air and not in vacuum, we
used air wavelengths to fit the position of the different lines. The next sections describe our approach in detail. 

\subsubsection{Multicomponent fit of emission and absorption lines}\label{sec:multicompVIMOS}

\begin{figure}[!htb]
\centering
\includegraphics[width=0.45\textwidth]{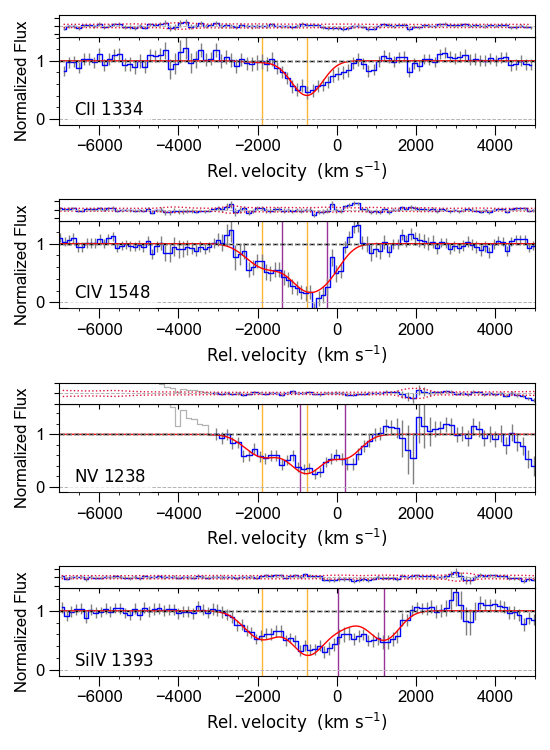}
\caption{VoigtFit models in velocity space. The normalised spectra, shown in blue with 3$\sigma$ error bars, display the regions in the vicinity of the absorption lines; the best-fit models are overlaid in red. Carbon, nitrogen and silicon lines are fitted simultaneously, with two kinematic components; the vertical lines identify the velocity centroids for each kinematic component. For the doublets, the reported velocities are calculated with respect to the systemic of the blue line of the doublet; the kinematic components associated with the blue transitions are in orange, those of the red transitions are in purple. The narrow top panels show the best-fit residuals, with 3$\sigma$ confidence intervals. Emission line contributions were removed prior to the VoigtFit modelling.
}\label{fig:Voigtfit}
\end{figure}

To model the low- and high-ionisation lines in the VIMOS spectrum, we used independent fits. This approach accounts for the fact that these species can trace different gas phases, potentially resulting in different systemic velocities and line widths (e.g. \citealt{Shen2011, Lanzuisi2015,Lanzuisi2024}). We classified the transitions based on their ionisation potential (IP). High-ionization lines (HILs) were those with an IP between 30 and 80 eV, and include: \nv $\lambda\lambda1238.82,1242.80$, \siiv $\lambda\lambda1393.76, 1402.77$, [\niv $\lambda1483.32$, \niv $\lambda1486.50$, \civ $\lambda\lambda1548.19, 1550.77$, and \heii $\lambda1640.42$. 
Low-ionization lines (LILs), with IP~$<30$~eV, included the following transitions: \cii $\lambda1334.53$, \siuii $\lambda1260.42$, \siuiii $\lambda\lambda1298.9, 1303.32$, 
the fluorescent doublets C\,{\sc{ii}}* $\lambda\lambda1335.66,1335.71$ and Si\,{\sc{ii}}* $\lambda\lambda1264.74,1265.02$, and the doublet \ciii $\lambda\lambda1906.68,1908.73$.  
We used single Gaussian profiles to fit the emission lines, and Voigt profiles to model the absorption line features of both  HILs (\nv, \civ, and \siiv) and LILs (\cii, C\,{\sc{ii}}*, Si\,{\sc{ii}}*, \siuiii). The fitting process used the Levenberg–Markwardt least-squares fitting code CAP-MPFIT (\citealt{Cappellari2017}). 

To achieve better constraints on the properties of transitions with lower S/N and to reduce fit degeneracy, especially in carbon and silicon fine structure lines with potential emission and absorption contributions, we simultaneously fitted transitions within the two individual groups of HILs and LILs (e.g. \citealt{Perna2015a}).  Namely, we constrained the wavelength separation between line transitions in accordance with atomic physics (considering their air wavelengths); moreover, we fixed their widths (in \kms) to be the same for all emission lines. For the emission line doublets \civ, \nv, and \siiv, the flux ratios were allowed to vary between the optically thick (f(B)/ f(R) = 1) and thin (f(B)/ f(R) = 2) limits, where R and B represent the red and blue transitions of each specific doublet (e.g. \citealt{DelZanna2002}). The line ratio \ciii f(1907)/f(1909), sensitive to the electron density, was allowed to vary within the physical variation range $0-1.6$ (e.g. \citealt{Keenan1992}). 
The HIL features show multiple peaks in absorption; therefore, we considered up to two kinematic components to model them. For the LILs in absorption, only one kinematic component was considered. During the fitting procedure, the flux ratio between the absorption lines of each species was treated as free parameter.


Figure \ref{fig:VIMOS_EM} shows the best-fit results for both HILs and LILs. The emission lines have a consistent systemic velocity  ($\Delta v_{LIL} = -3_{-6}^{+27}$~\kms; $\Delta v_{HIL} = 35_{-40}^{+5}$~\kms). However, the emission line profiles are broader in HILs ($\sigma_{HIL} = 630_{-5}^{+40}$~\kms) than in LILs ($\sigma_{LIL} = 330_{-5}^{+60}$~\kms). Similarly, HILs in absorption are broader and require two kinematic components: the first one is blueshifted by $\sim -900$~\kms, the second by $\sim -2000$~\kms; both have a Doppler parameter $b \sim 400$~\kms. Conversely, LILs in absorption only require a Voigt kinematic component, blueshifted by $\sim -900$~\kms and with $b \sim 400$~\kms; therefore, this kinematic component perfectly match the low-velocity one identified in the HILs in absorption. These properties allow us to classify GS133 as a mini-BAL AGN. 

As mentioned above, the degeneracy between emission and absorption contributions may affect our results. In particular, \nv, \siuii,  \siiv, \civ (and \cii) show complex P Cygni profiles and are therefore associated with two (three) line transitions with potential contributions both in absorption and emission. Our fitting approach allowed us to reduce this degeneracy in the specific species thanks to the simultaneous modelling of other transitions like \siuiii, detected only in absorption, and \heii and \ciii, detected only in emission.  
Nevertheless, a notable limitation is that our model does not identify any absorbing kinematic component at the systemic redshift, although such a component might exist. The combined impact of fit degeneracies and the low spectral resolution of the VIMOS data significantly constrains our ability to accurately model these complex line profiles. These limitations are further explored in the following sections.

\subsubsection{Modelling of absorption lines with VoigtFit}

After removing the emission line contributions inferred from the previous step, we modelled the most prominent absorption lines in the VIMOS spectrum with the Python package VoigtFit (\citealt{Krogager2018}). Namely, we considered \cii $\lambda 1334.53$, and the \nv $\lambda\lambda1238.82,1242.80$, \siiv $\lambda\lambda1393.76, 1402.77$, and \civ $\lambda\lambda1548.19, 1550.77$ doublets.

We fitted the transitions from various ions and elements simultaneously, without separating the HILs and LILs, considering two distinct kinematic components. This is justified by the fact that independent fit results presented in Sect. \ref{sec:multicompVIMOS} already showed that the high and low ionisation transitions have similar profiles at the lowest velocities. Figure \ref{fig:Voigtfit} shows the best-fit results for the four species. The LIL \cii is blue-shifted by $-778\pm 26$~\kms, and is symmetric, with $b = 360\pm 45$~\kms. The same blue-shifted kinematic component is also detected in the three HILs; these lines, however, also required a kinematic component with more extreme properties, namely $\Delta v~=~-1910\pm 50$~\kms and $b = 490\pm 80$~\kms.     
VoigtFit also provides a measurement of the column density for each transition and kinematic components; all best-fit parameters are reported in Table \ref{tab:UVproperties}.

\begin{table}[b]
\tabcolsep 3.2pt 
\centering
\caption{VoigtFit best-fit parameters and inferred column densities.}%
\begin{tabular}{lccc}
\hline

  &  log (N/cm$^{-2}$) & log (N$_{\rm H}$/cm$^{-2}$) & log (N$_{\rm H}^{c}$/cm$^{-2}$)\\
  & {\footnotesize (1)} & {\footnotesize (2)}& {\footnotesize (3)}\\
\hline
\hline
\multicolumn{3} {l} {$\Delta v = -778 \pm 26$~\kms, $b = 340\pm 50$~\kms}\\
\hline
\cii 1335& $15.23\pm 0.07$ & $>18.77$ & $>$18.9--19.0\\
\civ 1548, 1551 & $ 15.22\pm 0.05$ & $>18.76$ & $>$18.8--18.9\\
\nv 1239, 1243 & $15.32\pm 0.07$ & $>19.42$ & $>$19.53\\
\siiv 1394, 1403 & $14.88\pm 0.05$ & $>19.27$ & $>$19.5-20.6 \\

\hline
\hline
\multicolumn{3} {l} {$\Delta v = -1910 \pm 50$~\kms, $b = 410\pm 80$~\kms}\\
\hline
\cii 1335& $14.00\pm 0.70$ & --& --\\
\civ 1548, 1551 & $ 14.74\pm 0.07$ & $>$18.28 & $>$18.4--18.5\\
\nv 1239, 1243 & $14.99\pm 0.09$ & $>$19.09 & $>$19.2\\
\siiv 1394, 1403 & $14.60\pm 0.06$ & $>$18.99 & $>$19.2--20.3\\

\hline
\end{tabular} 
\tablefoot{
Col (1): Column density for individual species. Col (2): Inferred hydrogen column density lower limit from individual species. Col (3): Inferred hydrogen column density  lower limit corrected for dust depletion.
The \cii 1335 is well fitted with a single kinematic component (the one at $\Delta v \sim -800$~\kms); however, in the table we also report the tentative column density obtained by VoigtFit for the second kinematic component.   
}
\label{tab:UVproperties}
\end{table}

To estimate a conservative lower limit to hydrogen column density N$_{\rm H}$ in the outflow, we assumed no ionisation correction (e.g. N(C) = N(\civ)), and no correction for saturation. We also assumed solar abundances and the possible effects of depletion onto dust from \citet{Jenkins2009}, 
following \citet{Bordoloi2013}. In Table \ref{tab:UVproperties} we report for each transition a value without correction for dust depletion (third column) and a range of values for depletion corrected densities (last column) obtained considering the range of values in \citet[][see their Table 4]{Jenkins2009}. 

Overall, the column density lower limits obtained from the different species are within a narrow range of log~(N$_{\rm H}$/cm$^{-2}$)~$=18.5-19.5$, for both kinematic components. The main discrepancies are observed in the depletion corrected values for the \siiv transitions, which reach values up to log~(N$_{\rm H}$/cm$^{-2}$)~$=20.6$: the silicon element, however, has the largest and likely more uncertain depletion corrections, in the range [$-0.22, -1.36$]. Because of this, we considered in the derivation of the outflow energetics (Sect. \ref{sec:energetics}) an order of magnitude approximation log~(N$_{\rm H}$/cm$^{-2}$)~$= 19$. This column density is consistent with values generally associated with mini-BALs and NALs (e.g. \citealt{Laha2021}).

\subsection{Possible Balmer absorption in the NIRSpec spectrum}

\begin{figure}[!t]
\centering
\includegraphics[width=0.5\textwidth]{{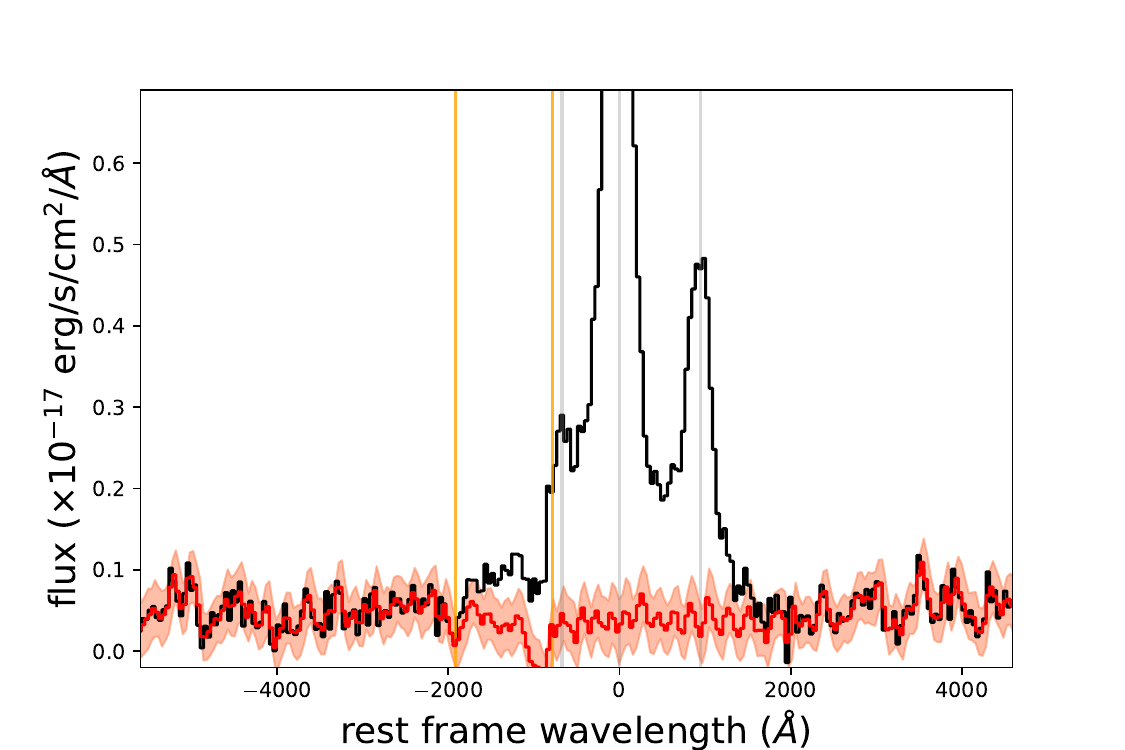}}

\caption{GS133 spectrum showing tentative Balmer absorption detection in the vicinity of \ha and \nii emission lines. The black curve shows the integrated spectrum in velocity space, with $v=0$~\kms at the position of the \ha line. The red curve (with 1$\sigma$ uncertainties in light-red) displays the residuals obtained by subtracting the best-fit Gaussian emission lines presented in Fig.~\ref{fig:opticalintegratedspectrum}. Two low S/N absorption features are detected close to the orange vertical lines, which mark the velocities of the mini-BAL features observed in the UV (see Table \ref{tab:UVproperties}). }\label{fig:balmerabsorption}
\end{figure}

\begin{figure*}[!htb]
\centering
\includegraphics[width=\textwidth]{{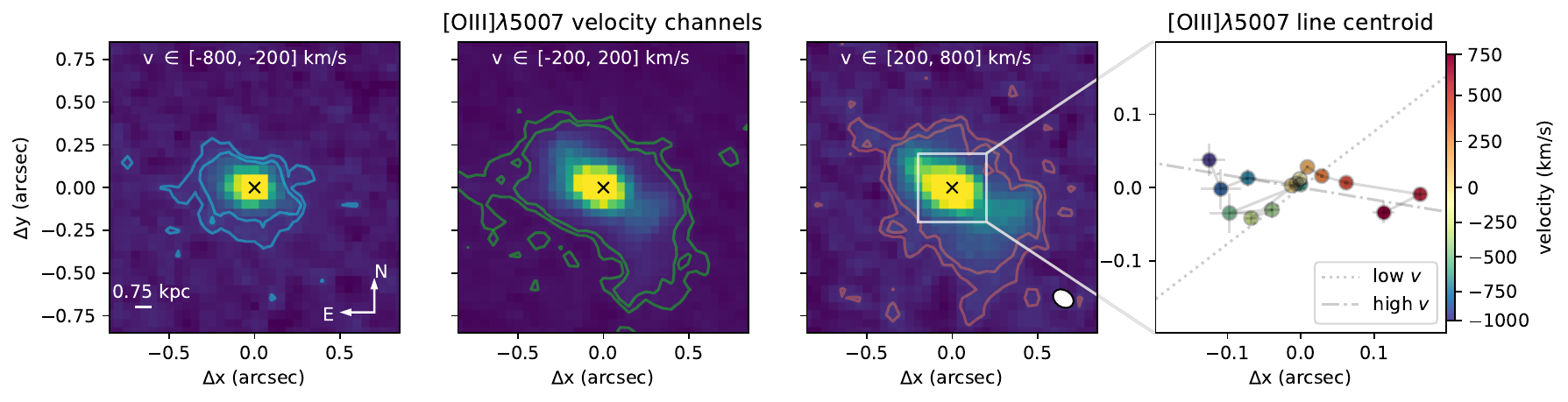}}

\caption{\oiii velocity-channel maps (first-to-third panels) and spectroastrometry \oiii 
line positions (right panel). First to third panels: The velocity-channel maps were extracted from the ranges labeled in the individual panels. For each flux distribution, we report 3 and $5\sigma$ contour levels. The black cross identifies the GS133 nucleus, as inferred from spectroastrometry position of spectrally integrated \oiii profile (i.e. considering all velocity channels together); the white ellipse in the third panel represents the NIRSpec point spread function (PSF) at $\sim 2.2$~$\mu$m (the \oiii observed wavelength) estimated by \citet{DEugenio2023}; a scale bar and a compass are also reported in the first map. Right panel: Spectroastrometry of the \oiii emission, i.e. centroid positions of the line in the different velocity
channels, covering a relatively small portion of the spatial extent shown in the velocity-channel maps (see white box in the third panel). The points are colour-coded by the channel velocity. The dot-dashed line roughly identifies the outflow axis, while the dotted line shows the tentative kinematic major axis of the GS133 rotating disc. 
}\label{fig:spectroastrometry}
\end{figure*}

The spatially integrated optical spectrum of GS133 in Fig. \ref{fig:opticalintegratedspectrum} displays two absorption features on the left side of the \ha-\nii complex. Figure \ref{fig:balmerabsorption} shows that they are at $\approx -1000$ and $\sim -2000$~\kms from the \ha peak, hence broadly matching the mini-BAL features observed in the VIMOS spectrum (see Fig. \ref{fig:Voigtfit}). Therefore, these two features could be interpreted as Balmer absorption, similar to those observed in a few broad-line AGN up to $z\sim 2$ in the pre-JWST era (e.g. \citealt{Hutchings2002, Schulze2018,Hamann2019}) and many JWST-selected Type 1 AGN (e.g. \citealt{Juodvzbalis2024, Kocevski2024,Matthee2024,  Wang2024arXiv240302304W}).

These Balmer absorption lines are thought to originate around the edge of the obscuring torus, that is eroded and accelerated by nuclear winds; in fact, their inferred distances from the central engine are larger than the size of the BLR but smaller than the size of the dusty torus based on detailed photoionization models using \textsc{Cloudy} (e.g. \citealt{Zhang2015ApJ...815..113Z, Shi2016,Juodvzbalis2024}). 
Alternatively, Balmer absorption could be associated with larger distances from the SMBH if the absorption is weak and the hydrogen density and N$_{\rm H}$ are low (see Sect. \ref{sec:cloudymodels}).

However, it is important to note that the absorption features detected in the integrated spectrum of GS133 have relatively low S/N ($\sim 2$). The subtraction of emission line contributions resulted in negative flux values in some velocity channels, indicating that the absorption feature shapes may be influenced by fit degeneracies. 
Consequently, we did not pursue further analysis of these tentative Balmer absorption detections.

\section{Spatially resolved analysis}\label{sec:spatiallyresolveddata}

This section presents the spatially resolved properties of the ionised gas in GS133 as inferred from the NIRSpec IFS data-cube.

\subsection{Velocity channels and spectro-astrometry}\label{sec:spectroastrometry}
Before performing a spaxel-by-spaxel spectral fit analysis, we generated \oiii velocity channel maps and used spectro-astrometry analysis to understand if the ionised gas is spatially resolved in our NIRSpec data.  

Figure \ref{fig:spectroastrometry} shows the \oiii maps obtained by integrating over three velocity ranges: [$-800, -200$]~\kms to cover the high-velocity blueshifted (approaching) component, [$-200, 200$]~\kms to cover the systemic emission, and [200, 800]~\kms  for the high-velocity redshifted (receding) gas. These maps show an extended morphology covering $\sim 1\arcsec$ (7.5~kpc) along the NE to SW direction, in particular in the systemic and redshifted components; the blueshifted emission is more compact (contours identify $3\sigma$ and $5\sigma$ levels, in each panel). 
The NIRSpec PSF (reported in the bottom-right part of the third map) is elongated along the same NE-SW direction, roughly corresponding to the direction of the NIRSpec IFS slices ($\sim 60^\circ$), but on smaller scales: FWHM $\sim 0.13\arcsec$ along the slices and $\sim 0.1\arcsec$ across the slices (\citealt{DEugenio2023}). Therefore, the \oiii emission in GS133 is spatially resolved. 

To better understand whether the high-velocity components are spatially separated from the systemic gas in the host galaxy, we also performed a spectro-astrometry analysis. This analysis consists in determining how the centroid position of the \oiii emission changes as a function
of velocity. We followed a methodology similar to that of  \citet{PereiraSantaella2018} and \citet{Lamperti2022}. We binned together the velocity channels in intervals of 120~\kms to increase the S/N, necessary to reliably determine the position of the peak of the emission. Then, we performed a fit with the Photutils package\footnote{\url{https://photutils.readthedocs.io/en/stable/}} to determine the peak position in each binned channel. In this way, we could determine the peak positions at sub-pixel scales. 

The \oiii spectro-astrometry positions are reported in the right panel of Fig. \ref{fig:spectroastrometry}. The low-velocity ($|v| \leq 120$~\kms) channels appear oriented along the NW to SE direction, while higher velocity ($|v|> 120$~\kms) are roughly perpendicular. This configuration suggests that GS133 hosts a rotating disk with a major axis perpendicular to the extended outflow (see also e.g. Fig. 3 in \citealt{Lamperti2022} for a similar configuration in a nearby galaxy). The orientation of the \oiii morphological major axis, which aligns with the outflow rather than the rotational axis, suggests that outflow emission is dominant over the unperturbed gas and that a significant portion of the ejected \oiii may has relatively low projected velocities. This kinematic configuration is further investigated in the next sections.

\begin{figure}[!htb]
\centering
\includegraphics[width=0.49\textwidth]{{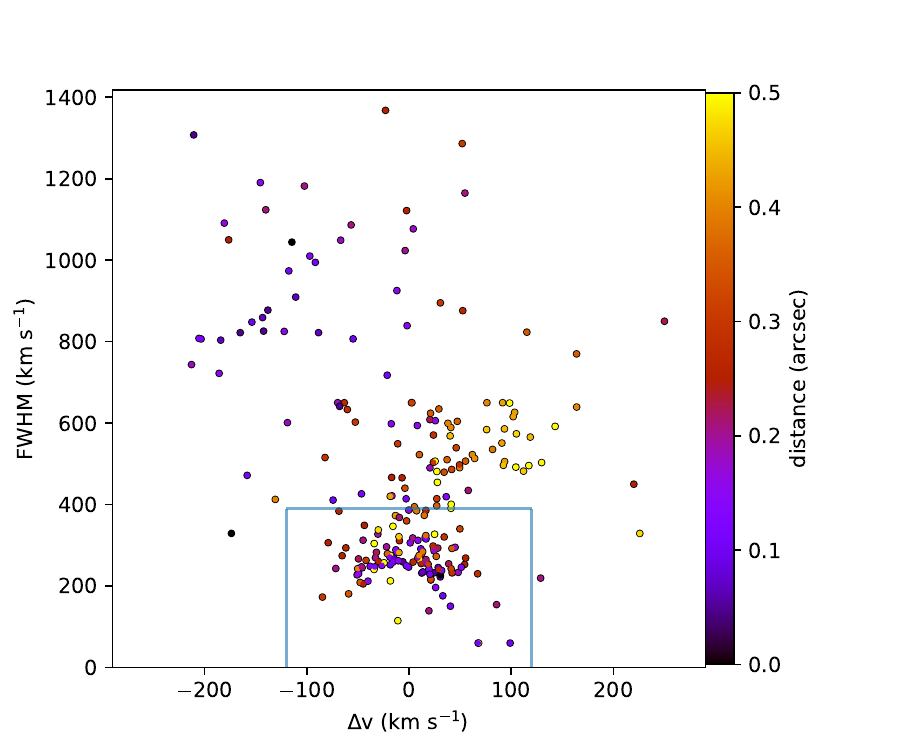}}

\caption{ Velocity diagram for the individual Gaussian components used to model the emission line profiles in the GS133 datacube. The measurements are coloured by the distance from the AGN. The blue lines isolate the Gaussian components with FWHM $< 400$~\kms and $|\Delta v|<120$~\kms used to separate the narrow emission line kinematics (reported in Fig. \ref{fig:diskmap}) from the outflow kinematics (Fig. \ref{fig:outflowmap}).
}\label{fig:DvFWHMdiagram}
\end{figure}

\begin{figure*}[!htb]
\centering
\includegraphics[width=0.8\textwidth]{{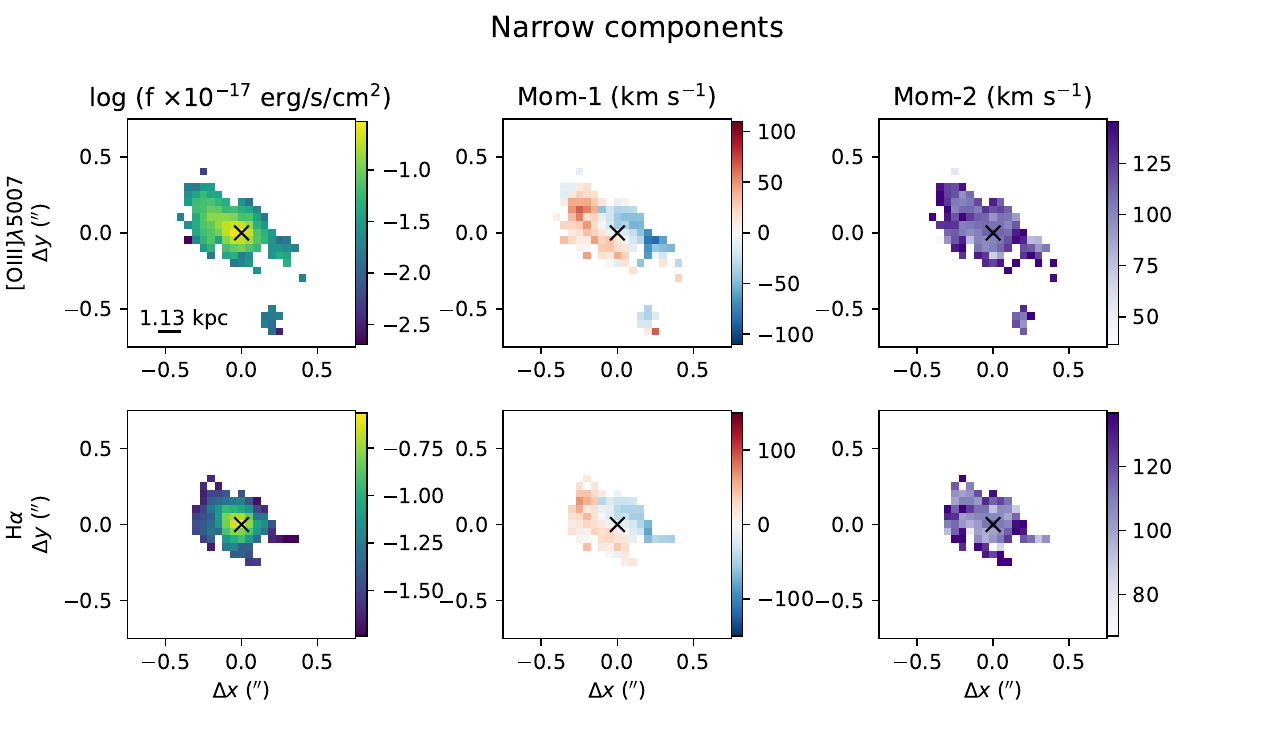}}

\caption{\oiii (top) and \ha (bottom) flux, moment-1 and moment-2 maps for the systemic kinematic component. A S/N cut of 4 has been applied to generate the maps. The moment-1 maps of both lines show evidence of rotating gas in the QSO host, but neither moment-2 map shows a peak at the nuclear position (identified by a black cross).
}\label{fig:diskmap}
\end{figure*}

\begin{figure*}[!htb]
\centering
\includegraphics[width=0.8\textwidth]{{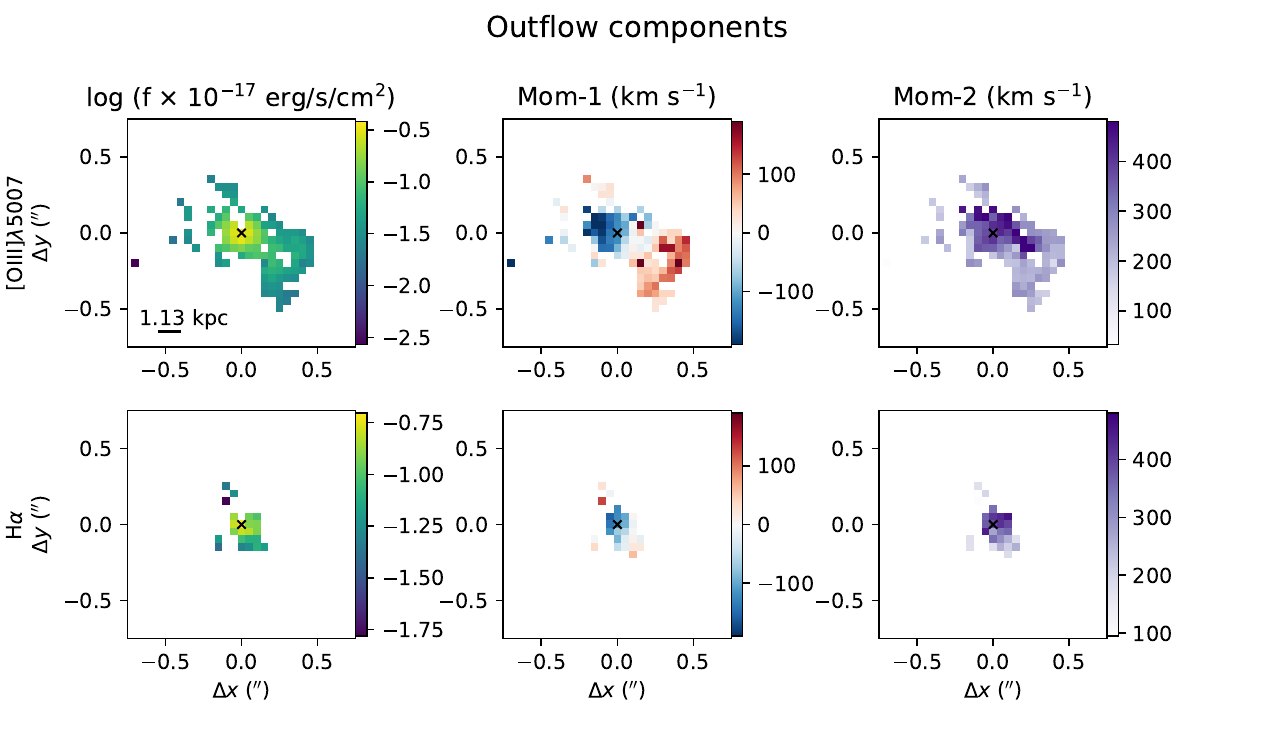}}

\caption{\oiii (top) and \ha (bottom) flux, moment-1 and moment-2 maps for the outflow kinematic component.
A S/N cut of 4 has been applied to generate the maps. The \oiii maps show a typical biconical outflow structure, oriented along the NE-SW direction, consistent with our spectro-astrometry analysis (Sect. \ref{sec:spectroastrometry}). The \ha flux distribution is less extended, as this Balmer line is fainter than \oiii. }\label{fig:outflowmap}
\end{figure*}

\subsection{Spaxel-by-spaxel multi-Gaussian fit}

To derive spatially resolved kinematic and physical properties of ionised gas, we fit the spectra of individual spaxels using the prescriptions already presented in Sect. \ref{sec:spatiallyintegratedNIRSpecdata}. We applied the BIC selection to determine where a multiple-Gaussian fit is required to statistically improve the best-fit model. For each Gaussian component included in the fit, a S/N~$>3$ was required. 
This approach ensures that the more complex, and potentially degenerate, multiple-component fits are used only where statistically justified. In most spaxels, two Gaussian components suffice, but a few spaxels closer to the AGN position require up to four Gaussian components.

Figure \ref{fig:DvFWHMdiagram} shows the $\Delta v-{\rm FWHM}$ velocity diagram obtained from our multi-Gaussian models, with the kinematic parameters of all Gaussian components required to fit the datacube.
The measurements are coloured by the distance from the nucleus. Although the diagram does not show a clear trend, we note that the highest FWHMs ($>400$~\kms) are sometimes associated with significant velocity offsets ($|\Delta v| > 150$~\kms), as observed in systems hosting AGN outflows (e.g. \citealt{Woo2016, Perna2022, Perna2023}). In particular, the most external regions (yellow points in the figure) show a positive trend, where both velocity and FWHM increase together; the broadest components are instead blueshifted and closer to the nucleus.  In contrast, the narrower Gaussian components have relatively small offsets from the zero velocity.

To check for the presence of rotationally supported motions (see the right panel in Fig. \ref{fig:spectroastrometry}), we selected in the $\Delta v-{\rm FWHM}$ plane the components with $|\Delta v|<120$~\kms and FWHM~$< 400$~\kms (i.e. within the blue box in Fig. \ref{fig:DvFWHMdiagram}), and constructed new datacubes containing only the systemic \ha and \oiii emission (see e.g. \citealt{Perna2022} for a similar approach). The flux distribution as well as the velocity offset and dispersion of the narrow \ha and \oiii lines are reported in Fig. \ref{fig:diskmap}. The $\Delta v$ maps display a regular pattern along the NW-SE direction, consistent with the spectro-astrometry analysis results (Sect. \ref{sec:spectroastrometry}), with a peak-to-peak velocity shift of $\sim 80$~\kms. The velocity dispersion maps do not display a clear peak at the position of the nucleus, as expected for a rotating disk (e.g. \citealt{ForsterSchreiber2018}). This could be due to the presence of the outflow and fit degeneracies. 
Indeed, the $\Delta v$ map shows some redshifted emission towards NE and blueshifted towards SW (mostly in \oiii), which do not follow the rotation pattern seen in the more central regions; this component could be similarly due to a portion of the ejected gas with relatively small projected velocities. This scenario is further discussed in Sect. \ref{sec:geometry}.

Figure \ref{fig:outflowmap} displays the flux, velocity offset, and velocity dispersion obtained for the outflow Gaussian components (outside the box identified in Fig. \ref{fig:DvFWHMdiagram}). The \oiii gas is more extended than \ha, and shows a bi-conical geometry, typical of AGN outflows (e.g. \citealt{Venturi2018, Perna2022, Falcone2024}). The ionised gas is blueshifted ($\Delta v\sim -280$~\kms) towards NE, and redshifted ($\Delta v\sim 100$~\kms) towards SW, up to $\sim 3$~kpc. The line widths reach values as high as $\sim 450$~\kms in the circum-nuclear regions, in line with the integrated fit results (Sect. \ref{sec:spatiallyintegratedNIRSpecdata}). 

For completeness, in Fig.~\ref{fig:totalprofilemaps} we report the flux and velocity maps obtained from the total, integrated emission lines, hence without separating the systemic from the outflow components. In these maps, the signatures of the rotating disk and the bi-conical outflow configuration are less pronounced, underlining the complex kinematics in GS133.

\section{Multi-wavelength diagnostics}\label{sec:diagnostics}

\subsection{AGN classification with optical line ratio diagnostics}\label{sec:bpt}

\begin{figure}[!t]
\centering
\includegraphics[width=0.5\textwidth]{{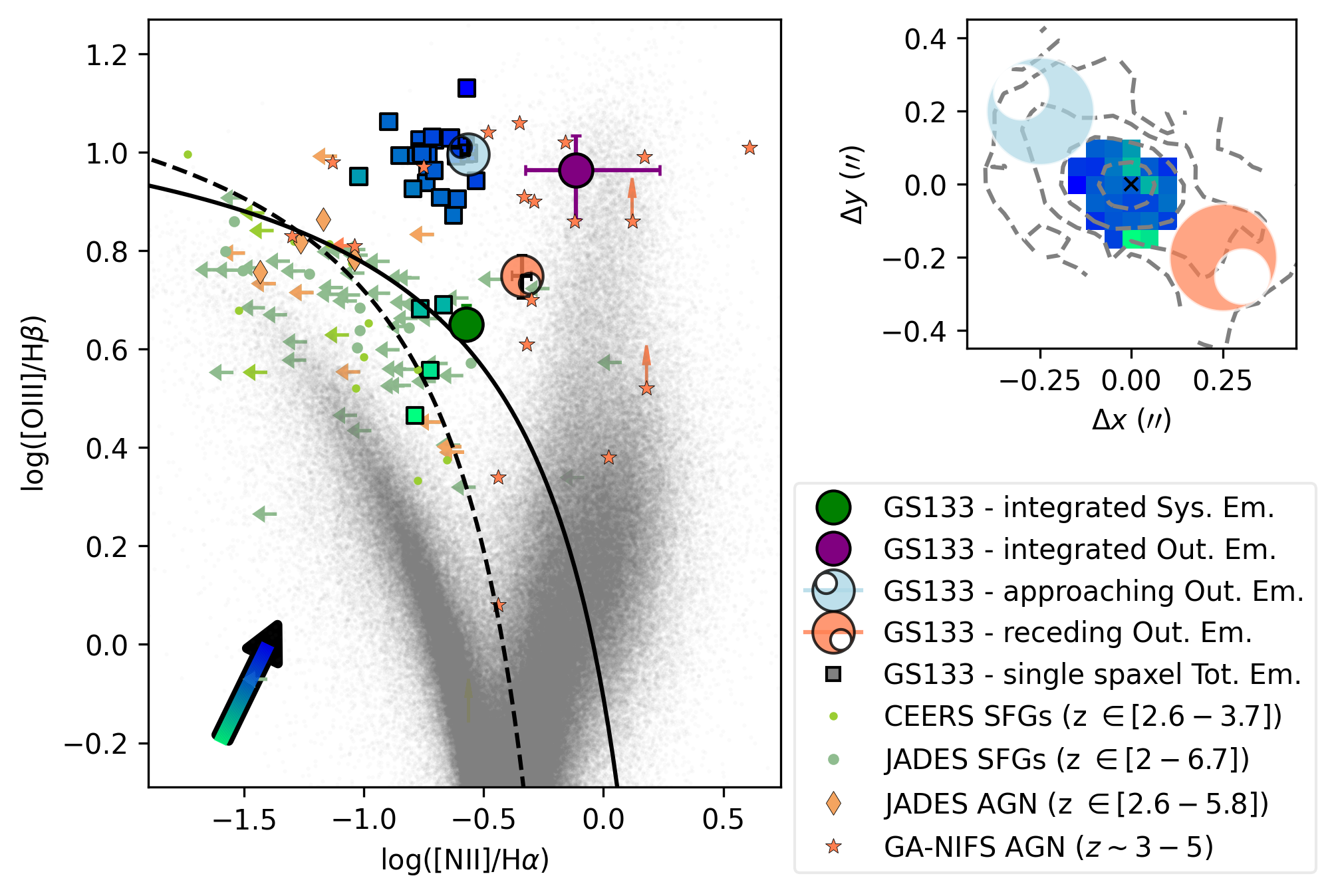}}

\caption{Standard BPT diagnostic diagram. The colour-coded squares show GS133 single-spaxel measurements associated with the spatial regions shown in the top-right panel; green-to-blue colours mark increasing line ratios, as indicated with the arrow in the bottom-left part of the diagram; only spaxels with S/N$~>3$ in all lines are shown. The large green and purple circles display the line ratios measured from $r=0.5$\arcsec\ spatially integrated regions, for the systemic and outflow components, as labelled: the large crescent moon symbols refer to the outflow component line ratios measured in circular regions marked in the inset (with the same crescent moon symbols). The solid (\citealt{Kewley2001}) and dashed (\citealt{Kauffmann2003}) curves commonly used to separate purely star-forming galaxies (below the curves) from AGN (above the curves) are also reported. Finally, the BPT shows local galaxies from SDSS (\citealt{Abazajian2009}, indicated in grey), and additional measurements for $z = 2.6-6.7$ sources from the literature: star-forming galaxies are marked with small green symbols and refer to CEERS (\citealt{Calabro2023}) and JADES (\citealt{Scholtz2023}) surveys, while AGN are marked with small orange symbols and refer to JADES (\citealt{Scholtz2023}) and GA-NIFS (\citealt{Perna2023b}) surveys.}\label{fig:BPT}
\end{figure}

We investigated the dominant ionisation source for the emitting gas across the GS133 host using
the classical ``Baldwin, Phillips \& Terlevich'' (BPT) diagram (\citealt{Baldwin1981}). 
Figure \ref{fig:BPT} shows the distribution of the flux ratio diagnostics across the GS133 host galaxy extension (green-to-blue squares for increasing x- and y-axis values of the BPT diagram). The figure also shows a few spatially integrated flux ratios: the green and purple large circles refer to the systemic and outflow components obtained from the spectrum in Fig. \ref{fig:opticalintegratedspectrum}; the light-blue and light-red crescent moon markers display the flux ratios of the outflow component in the integrated spectra extracted from the approaching and receding parts of the biconical outflow, respectively (using circular apertures with $r=0.15$\arcsec, see inset in Fig.~\ref{fig:BPT}), and reported in Figs. \ref{fig:NEcone} and \ref{fig:SWcone}. 
For reference, the BPT also displays other optical line ratio measurements from the literature, for low-$z$ SDSS galaxies (small grey points), and  star-forming galaxies and AGN at $z > 2.6$ recently observed by JWST (from \citealt{Scholtz2023,Perna2023b,Calabro2023}), as well as the  demarcation lines used to separate galaxies and AGN at $z\sim 0$ from \citet{Kewley2001} and \citet{kauffmann_2003}. 

The spatially resolved line ratio measurements do not show a significant variation across the GS133 host galaxy extension, although a tentative trend can be observed. The lowest \oiii/\hb and \nii/\ha line ratios are observed in the outskirts of the galaxy, towards NW and SE (see the map in the inset), and are very close to the BPT locus of star-forming galaxies at high-$z$ (green symbols in the figure).
Instead, the highest line ratio measurements are along the outflow axis, likely corresponding to the AGN ionisation cones of GS133. In fact, these high line ratios are in a BPT region free from contamination by star forming galaxies at any redshift, and only populated by AGN systems (e.g. \citealt{Scholtz2023, Perna2023,Perna2023b, Parlanti2024a,Decarli2024}).

The spatially integrated measurements reported in the figure (large circles and crescent moon symbols) similarly indicate that the systemic narrow emission is consistent with star-formation ionisation, while the outflow components are AGN dominated (see also line ratios reported in Table \ref{tab:OPTlineratios}).

In conclusion, both spatially integrated and spatially resolved line ratios support an AGN classification based on the standard BPT diagnostic diagram. In fact, while the BPT diagram may lose sensitivity in distinguishing between AGN and star-forming galaxies in low-mass, low-metallicity systems at high redshift, it remains effective for identifying AGN in more massive and metal-enriched galaxies (e.g. Z$> 0.5$~Z$_\odot$, \citealt{Feltre2016}).

\subsection{AGN classification with UV line ratio diagnostics}\label{sec:UVdiagnostics}

The direct observation of high-ionisation emission lines in the UV spectrum of GS133, such as N\,{\sc{v}}$\lambda\lambda$1239,43, C\,{\sc{iv}}$\lambda\lambda$1548,51, He\,{\sc{ii}}$\lambda$1640, also in combination with other UV transitions, can be used to obtain a further confirmation of the presence of AGN emission in this high-$z$ source (e.g. \citealt{Mascia2023, Scholtz2023, Maiolino2023b}). 

In Table \ref{tab:UVlineratios} we report the main UV line ratios commonly used to distinguish between AGN and star-forming galaxies in the literature (e.g. \citealt{Feltre2016, Nakajima2018, Perna2023b, Topping2024}). All of these line ratios are compatible with AGN ionisation, which is consistent with the X-ray and optical BPT classifications introduced in the previous sections.

\begin{table*}[h]
\tabcolsep 3.5pt 
\centering
\caption{Optical emission line ratios from NIRSpec integrated spectra.}%
\begin{tabular}{l|ccc|ccc|ccc}
\hline

Line ratio  & \multicolumn{3}{c}{Integ. Spec. ($r = 0.5$\arcsec)}&\multicolumn{3}{|c|}{NE cone ($r = 0.15$\arcsec)} & \multicolumn{3}{|c}{SW cone ($r = 0.15$\arcsec)}\\
        & tot & sys & out &  tot & sys & out &  tot & sys & out  \\

\hline
\hline

log(\oiii/\hb) & $0.66_{-0.02}^{+0.01}$ & $0.65_{-0.01}^{+0.04}$&  $0.96_{-0.10}^{0.07}$ & $0.96_{-0.02}^{+0.03}$ & $1.07_{-0.03}^{+0.10}$ & $0.82_{-0.03}^{+0.09}$ &   $0.84\pm 0.1$ & $1.01_{-0.03}^{+0.09}$ & $0.75_{-0.04}^{+0.01}$ \\
log(\nii/\ha) & $-0.57_{-0.03}^{+0.16}$ & $-0.57_{-0.01}^{+0.06}$ & $-0.11_{-0.21}^{+0.35}$ & $-0.54_{-0.04}^{+0.03}$ &$-0.45_{-0.17}^{+0.08}$ & $-0.40\pm 0.1$ & $-0.51\pm 0.01$ & $-0.84_{-0.05}^{0.24}$ & $-0.32_{-0.02}^{+0.04}$\\

\hline

\hline

\end{tabular} 
\tablefoot{Integrated spectrum emission lines from best-fit shown in Fig. \ref{fig:opticalintegratedspectrum}; NE and SW cones from best-fit shown in Figs. \ref{fig:NEcone} and \ref{fig:SWcone}, respectively.   
}  
\label{tab:OPTlineratios}
\end{table*}

\begin{table*}[h]
\tabcolsep 5pt 
\centering

\caption{UV emission line ratios from the VIMOS spectrum.}\label{tab:UVlineratios}%
\begin{tabular}{@{}ccccccccc@{}}
\hline
   $\frac{\rm CIII]}{\rm HeII}$ &    $\frac{\rm CIV} {\rm HeII}$ &   $\frac{\rm CIV}{\rm CIII]}$ &   $\frac{\rm CIII] + CIV}{\rm HeII}$ &$\frac{\rm NV}{\rm HeII}$ &$\frac{\rm CIV}{\rm NV}$ &$\frac{\rm OIII]}{\rm HeII}$ &   log(L(\heii) / [\ergs])  \\
\hline

$-0.32_{-0.02}^{0.18}$ & $0.35\pm 0.01$ & $0.68_{-0.06}^{+0.10}$ & $0.44_{-0.01}^{+0.04}$ & $0.14_{-0.01}^{+0.01}$ & $0.21_{-0.01}^{+0.02}$& $< -0.96$ & $42.2\pm 0.1$\\

\hline
\end{tabular}
\tablefoot{O\,{\sc{iii}}]$\lambda\lambda$1660.81,1666.15 is not detected in the VIMOS spectrum.}
\end{table*}

\subsection{CT AGN classification with L({\rm \oiii})/ L(X-ray) diagnostic}

X-ray diagnostics are the most reliable methods for identifying CT AGN (e.g. \citealt{Lanzuisi2017}). Previous analyses by \citet{Luo2017} and \citet{Li2019} of the Chandra X-ray spectrum for GS133 suggested column densities of N$_{\rm H}$~$\approx~1.5\times~10^{24}$~cm$^{-2}$,  
which is around the threshold typically used to classify AGN as CT. However, due to the significant uncertainties associated with the low number of X-ray counts typical for high-redshift sources like GS133, where only 70.7 aperture-corrected source counts are available in the 0.5–7 keV band (\citealt{Luo2017}), the true column density could be lower.
To strengthen the case, multi-wavelength diagnostics have been employed by \citet{Guo2021}. They compared the X-ray to mid-infrared (6~$\mu$m) luminosity, commonly used to identify CT sources as the infrared emission is less affected by obscuration than X-rays; they found obscuration exceeding $1.5\times 10^{24}$ cm$^{-2}$. In this section, we tested an additional diagnostic to further support the CT classification of the AGN in GS133.

\citet{Maiolino1998} provided a diagnostic based on the ratio between the observed 2--10~keV and the reddening-corrected \oiii luminosities to infer the column density of obscured AGN. Adopting their methodology, we derived a luminosity ratio log(${\rm L_{X}/L_{[OIII]} }$)~$= -1.4$, corresponding to a lower limit log(${\rm N_H / cm^{-2}}$)~$> 2\times 10^{24}$ (see their Fig. 6). Specifically, the X-ray to \oiii luminosity ratio was calculated using the \oiii outflow luminosity reported in Table \ref{tab:OPTproperties} (hence excluding any possible contribution from star formation). 
The \oiii luminosity was corrected for dust extinction considering an intrinsic Balmer decrement of 2.86 and assuming the \citet{Cardelli1989} extinction law. 
Therefore, this result further supports the CT nature of GS133.

\section{Dynamical mass}\label{sec:mdyn}

Having an estimation of the kinematics and extent of a galaxy allows us to provide a constraint on the dynamical
mass of the galaxy, M$_{\rm dyn}$. Under the assumption that the 
gas is distributed in a flat disk, the dynamical mass enclosed within a radius R is M$_{\rm dyn} = 2.33\times 10^5~v_{\rm circ}^2~$R~M$_\odot$, where $v_{\rm circ}$ is the circular velocity in \kms at a galactocentric distance R, given in kpc (\citealt{Walter2004, Neeleman2021, Perna2022}). Since we could only marginally resolve the disk structure in GS133, we assumed an extent of R~$=1$~kpc, typical of $z~= 3-4$ galaxies (\citealt{Kartaltepe2023, Costantin2024}), and that the FWHM (i.e. $2.355\times \sigma$) of the systemic \ha line can be used as a proxy for the circular velocity, so that $v_{\rm circ} = 0.75~{\rm FWHM} / {\rm sin}(i) = 300$~\kms, assuming a mid-range inclination of 55$^\circ$ following, for instance, \citet{Neeleman2021} and \citet{Decarli2018}. 

Under these assumptions, the dynamical mass of GS133 is M$_{\rm dyn} = 2\times 10^{10}$~M$_\odot$, broadly consistent with the stellar mass inferred by \citet{Guo2020} and Circosta et al. (in prep.) from spectral-energy-distribution (SED) analysis: M$_* = 5\times 10^{9}$~M$_\odot$, and $3\times 10^{10}$~M$_\odot$, respectively. The difference in stellar mass estimates can be explained by the common degeneracy between AGN and stellar components in the optical and UV regime (e.g., \citealt{Ciesla2015}).
A more comprehensive comparison of the SED fitting results will be provided in Circosta et al. (in prep.).

\section{Outflow properties}\label{sec:energetics}

In this section we measure the mass rate and energetics of the ionised outflow as inferred from the blueshifted outflow components of absorbing (\cii, \civ, and \nv) and emitting  (\hb and \oiii) line transitions. We begin with standard assumptions commonly used in the literature, then refine these measurements by applying detailed photoionisation modelling for the absorption lines and 3D kinematic modelling for the emission lines. We also compare our results with outflow model predictions and general AGN properties to exclude less plausible assumptions, aiming to derive more accurate outflow properties for both the absorbing and emitting gas.

\subsection{Energetics of absorbing gas outflow}\label{sec:cloudymodels}

We measured the mass rate of the absorbing outflow gas as inferred from the blueshifted mini-BAL components of \cii, \civ, and \nv. Following \citet{Bordoloi2013}, we assumed for simplicity a thin-shell approximation ($\Omega =  4\pi$), and maximised to unity the angular and clumpiness covering factors ($C_\Omega C_f =1$): 

\begin{equation}\label{eq:Mdot}
    \dot M (Abs) \simeq  \frac{N_{\rm H}}{10^{20}~{\rm cm^{-2}}} \frac{R_{\rm out}}{1~ \rm kpc} \frac{v_{\rm out}}{100~ {\rm km~s^{-1}}}~ {\rm M_\odot~yr^{-1}}
\end{equation}

where N$_{\rm H}$ is the hydrogen column density and $R_{\rm out}$ and $v_{\rm out}$ are the mini-BAL galactocentric radius and velocity, respectively. The kinetic power of the absorbing material can be derived from the relation $\dot E = 1/2 \dot M v^2$:

\begin{equation}\label{eq:Edot}
    \dot E (Abs) \simeq 5\times 10^{39} \frac{N_{\rm H}}{10^{20}~{\rm cm^{-2}}} \frac{R_{\rm out}}{1~ \rm kpc} \left( \frac{v_{\rm out}}{100~ {\rm km~s^{-1}}} \right)^3 ~{\rm erg~s^{-1}}
\end{equation}

(see also \citealt{Hamann2019}). We assumed that the absorbing outflow is related to the two distinct components travelling at different velocities
(Fig.~\ref{fig:Voigtfit}), and that $v_{\rm out}$ is approximated by the velocity shift $\Delta v$ of a given component, hence $\sim -800$~\kms and $\sim -1900$~\kms (Table \ref{tab:UVproperties}).
For the extent of the mini-BAL outflow, we initially used a range of values between 1~pc and 10~kpc (following \citealt{Dunn2010, Saturni2016, Moravec2017, Bruni2019}), as it cannot be inferred without knowing the hydrogen density in the absorbing ejected gas. 
For the column density estimate, we consider median values inferred from dust-depletion corrected N$_{\rm H}^c$ reported in Table \ref{tab:UVproperties}, excluding the \siiv estimates as they are associated with more uncertain corrections (see \citealt{Jenkins2009}): $8\times 10^{18}$~cm$^{-2}$ and $1\times 10^{19}$~cm$^{-2}$ for the low- and high-velocity outflows, respectively.
We obtained $\dot M(Abs) = 0.003-30$ M$_\odot$~yr$^{-1}$ and  $\dot E(Abs) = (0.0004-4)\times 10^{43}$~\ergs, considering the extremely large range of possible values for $R_{\rm out}$, and computing the sum of the contributions from two kinematic components (with $\sim 70\%$ coming from the faster and denser component). In the next subsection, these measurements will be refined taking into account detailed photoionisation models. 

\subsection{Photoionisation modelling for ejected absorbing gas}

\begin{figure*}[!htb]
\centering
\includegraphics[width=\columnwidth]{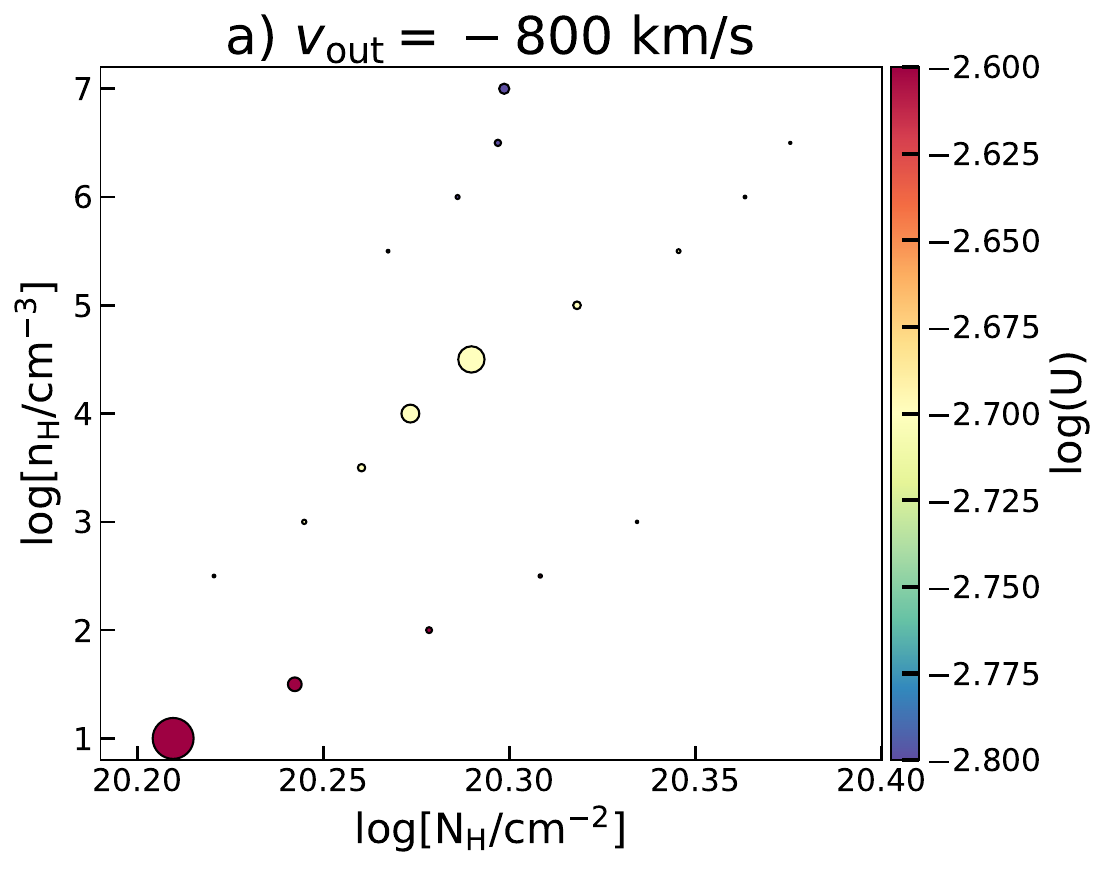}
\includegraphics[width=\columnwidth]{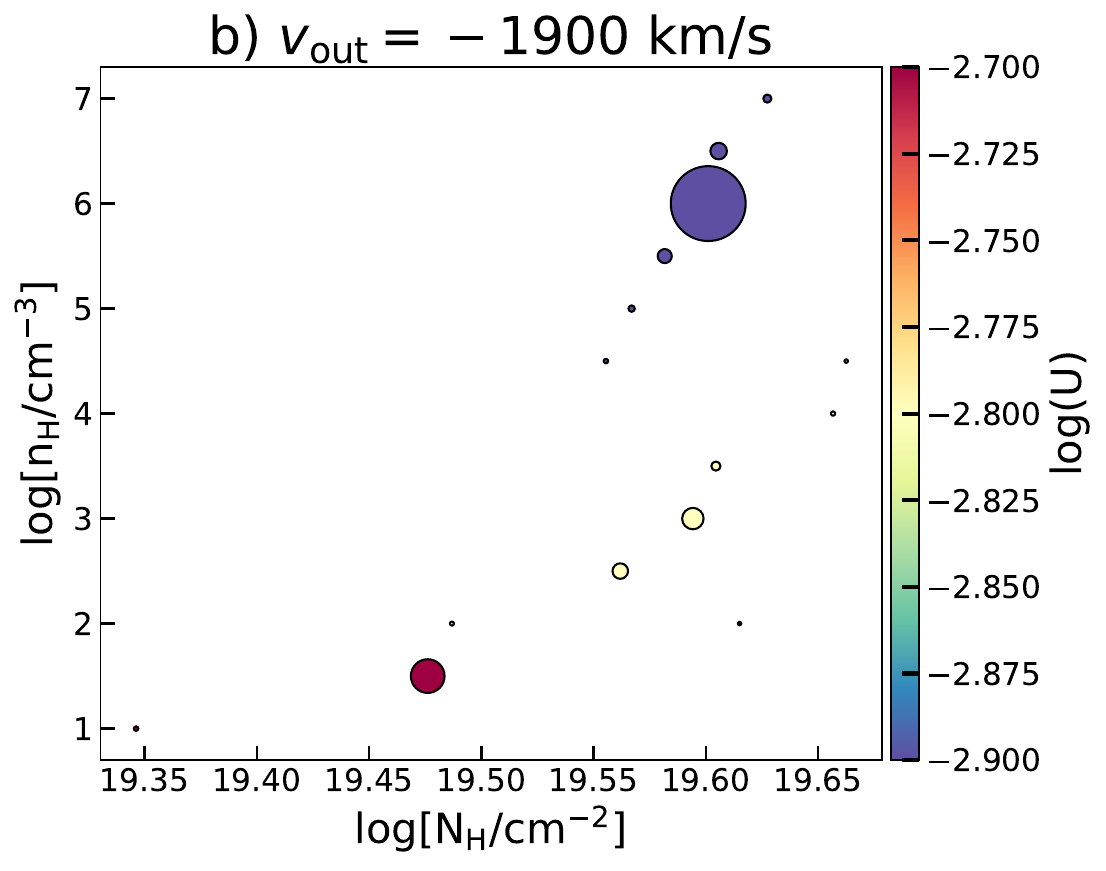}

\caption{Hydrogen densities and column densities predicted by \textsc{Cloudy} photoionisation models for the two mini-BAL components. Only models with $\chi^2 < 5$ are shown and the sizes of the symbols are inversely scaled with $\chi^2$. 
In addition, models are colour coded according to their ionisation parameters.
Both outflow components have low-density and high-density solutions, although the high-density solution would imply extremely thin layers of gas.}\label{fig:model_sol}
\end{figure*}

To infer the physical conditions within the mini-BAL, we computed a series of simulations using the photoionisation code \textsc{Cloudy} (c17.03; \citealt{Ferland2017}) and compared them with observations.
We set up a series of 1D single-cloud models with the ionising SED set to a \textsc{Cloudy} built-in multicomponent AGN spectrum described by the following formula
\begin{equation}
    F_\nu = \nu ^{\alpha _{\rm UV}}{\rm exp}(-h\nu /kT_{\rm BB}){\rm exp}(0.01~{\rm Ryd}/h\nu) + a\nu ^{\alpha _{\rm x}}.
\end{equation}
Specifically, we set the cut-off temperature for the accretion disk to $T_{\rm BB}=10^6$ K, the optical-to-X-ray ratio to $\alpha _{\rm ox} = -1.4$, the X-ray slope to $\alpha _{\rm x} = -1$, and the UV slope to $\alpha _{\rm uv} = -0.5$ (by adjusting parameter $a$ in the equation above).
The SED we adopted roughly reproduces the shape of the observed UV continuum observed by VLT/VIMOS.
For the absorbing clouds, we explored a wide range of ionisation parameters\footnote{Throughout this work $U$ is defined as the dimensionless ratio between the ionising photon flux and the product of hydrogen density and the speed of light.}, $U$, from $10^{-4.5}$ to $1$.
We set the clouds to be isobaric and vary the density in a range of $1 \leq \log(n_{\rm H}/{\rm cm^{-3}}) \leq 7$.
All models are computed until a cumulative hydrogen column density of $N_{\rm H} = 10^{22}~{\rm cm^{-2}}$, and cumulative column densities for different atomic/ionic transitions as functions of depth are saved individually for the search of solutions that match the derived values for GS133.
For the chemical abundances, we adopted the solar abundance set of \citet{grevesse_2010_solar} and set the ratio N/C as a free parameter to account for variations due to chemical enrichment history \citep{maiolino_2019}.
Finally, since Si is usually highly depleted onto dust grains in the ISM \citep[e.g.][]{Jenkins2009}, we set Si/C as an additional free parameter to probe the level of dust depletion.

We considered separate models for the two kinematic components at $-800$~\kms and $-1900$~\kms. We used a single-cloud model, where both \cii and \civ can be found, but \civ absorption happens in the more inner region of the cloud due to its higher ionization potential. This means that when \cii has reached a large column density at a certain depth in the cloud, the column density of \civ usually has already reached its maximum value.
Specifically, for the low-velocity component, we searched for models with column densities of the \cii that match our derivations from the analysis of the VIMOS spectrum. Then, we allowed the column density of \civ to differ from the measured value by $\pm 0.15$~dex (i.e. $3\sigma$, Table \ref{tab:UVproperties}), and scaled the N/C and Si/C abundances to reproduce the column densities measured for \nv and \siiv transitions.
We also performed a minimum $\chi^2$ search using the measured column densities for the above transitions and obtained consistent results.

In Fig.~\ref{fig:model_sol} we show the best-fit $n_{\rm H}$ and $N_{\rm H}$ for the absorbing gas.
For the mini-BAL component with $v_{\rm out}\sim -800$~\kms, we found low-density models with log~(N$_{\rm H}$/cm$^{-2}$) = 20.2--20.3, log~(n$_{\rm H}$/cm$^{-3}$) = 1--2, and log ($U$) in the range [-2.7, -2.6] can well reproduce our measurements.
High-density models with log~(n$_{\rm H}$/cm$^{-3}$) = 4--5 also have a good match with our measurements.
All these models require a high log(N/C) = 1 to reproduce \nv absorption, consistent with CNO equilibrium ratio in asymptotic giant branch (AGB) stars \citep{maeder_cno_2015}. However, since there is no strong nebular emission of \niv and \niii in the VIMOS spectrum, we caution the interpretation of the high N/C.
As an alternative explanation, due to the high ionisation potential of \nv, \nv absorption might actually occur in a separate density-bounded cloud closer to the central AGN, which cannot be described with single-cloud models; fit degeneracy between emission and absorption contributions could also explain the high N/C.
Based on the best-fit Si/C, we inferred a depletion factor of 0.52 dex for the \siiv. The depletion of \siiv is consistent with a moderate amount of dust depletion found in the compilation of MW sightlines by \citet{Jenkins2009} with a unified depletion strength of $F_{*}=0.25$.
In comparison, a stronger depletion strength of $F_{*}=0.5$ is typical for star-forming galaxies at $z \lesssim 0.1$ \citep{gunasekera_dustdepl_2023}.
For the high-velocity component, we found low-density models with log~(N$_{\rm H}$/cm$^{-2}$) = 19.4--19.6, log~(n$_{\rm H}$/cm$^{-3}$) = 1.5--3.0, and log ($U$) in the range [-2.8, -2.7] can reproduce our measurements if we leave N/C and Si/C as free parameters.
There is a high-density solution at log(n$_{\rm H}$/cm$^{-3}$) $\approx$ 6 as well. If we further constrain N/C and Si/C in the models to the values obtained for the slow component, the preferred models for the fast component would have log~(N$_{\rm H}$/cm$^{-2}$) $\approx$ 19.6, log~(n$_{\rm H}$/cm$^{-3}$) = 2.0--4.0, and log ($U$) in the range [-2.8, -2.7].

For both low- and high-velocity components, our low-density models suggest that the effective thickness of the absorbing gas (i.e., N$_{\rm H}$/n$_{\rm H}$) can span a range from 4 to 0.04~pc if it has a volume filling factor close to unity, consistent with  previous results for low-$z$ systems from the literature (e.g. \citealt{Choi2022}).
In comparison, high-density solutions correspond to physical scales of $6\times 10^{-4}-10^{-3}$ pc for the low-velocity component and $10^{-5}$ pc for the high-velocity component.

Knowing the hydrogen density and ionisation parameter for the cloud from \textsc{Cloudy} models, we could also infer the distance of the absorbing material from the SMBH, from 

\begin{equation}
    U = \frac{Q}{4 \pi R_{\rm out}^2 n_{\rm H} c},
\end{equation}

where $Q$ is the number of ionising photons per unit time, proportional to the AGN bolometric luminosity (see e.g. Eq. 4 in \citealt{Baron2019b}), and $c$ is the speed of light. By solving this equation for R$_{\rm out}$, we obtained distances of 1--10~kpc for the two kinematic components with low-density \textsc{Cloudy} models reported above. We note that much smaller distances would require significantly higher densities of n$_{\rm H}$ ($\approx 10^6$~cm$^{-3}$), which, however, would be associated with an unphysical thickness of the cloud, orders of magnitude smaller than typical dimensions of a single cloud in the BLR ($\lesssim 10^{-3}$~pc), and smaller than normally assumed in BAL models (e.g. r $\sim 0.3-3$~pc in \citealt{ Ishibashi2024}). 

Therefore, our \textsc{Cloudy} models suggest that the mini-BAL material in GS133 is at the same distance of emitting gas and not confined in the nuclear (parsec-scale) regions. 
While single-cloud models might be overly simplistic for a system like GS133, which exhibits multiple kinematic components in both low- and high-ionisation lines, as well as an extended, bi-conical \oiii outflow, more complex models with multiple clouds would introduce greater degeneracies (as they require the modelling of $>10$ line transitions; e.g. \citealt{Marconi2024}).
Therefore, our approach represents a compromise, providing indicative properties of the mini-BALs in GS133. 

Using the best-fit \textsc{Cloudy} parameters, we recalculated the UV outflow energetics. For the low-velocity component, we found mass outflow rates of
14 -- 140~\Msunyr and kinetic powers of (5 -- 50)~$\times 10^{42}$~\ergs, with the range reflecting an outflow extent from 1 to 10 kpc.  
For the high-velocity component, the corresponding values are 6 -- 60~\Msunyr and (1 -- 10)~$\times 10^{43}$~\ergs, with the same range for the outflow extent.

\begin{figure*}[!t]
\centering
\includegraphics[width=0.95\textwidth]{{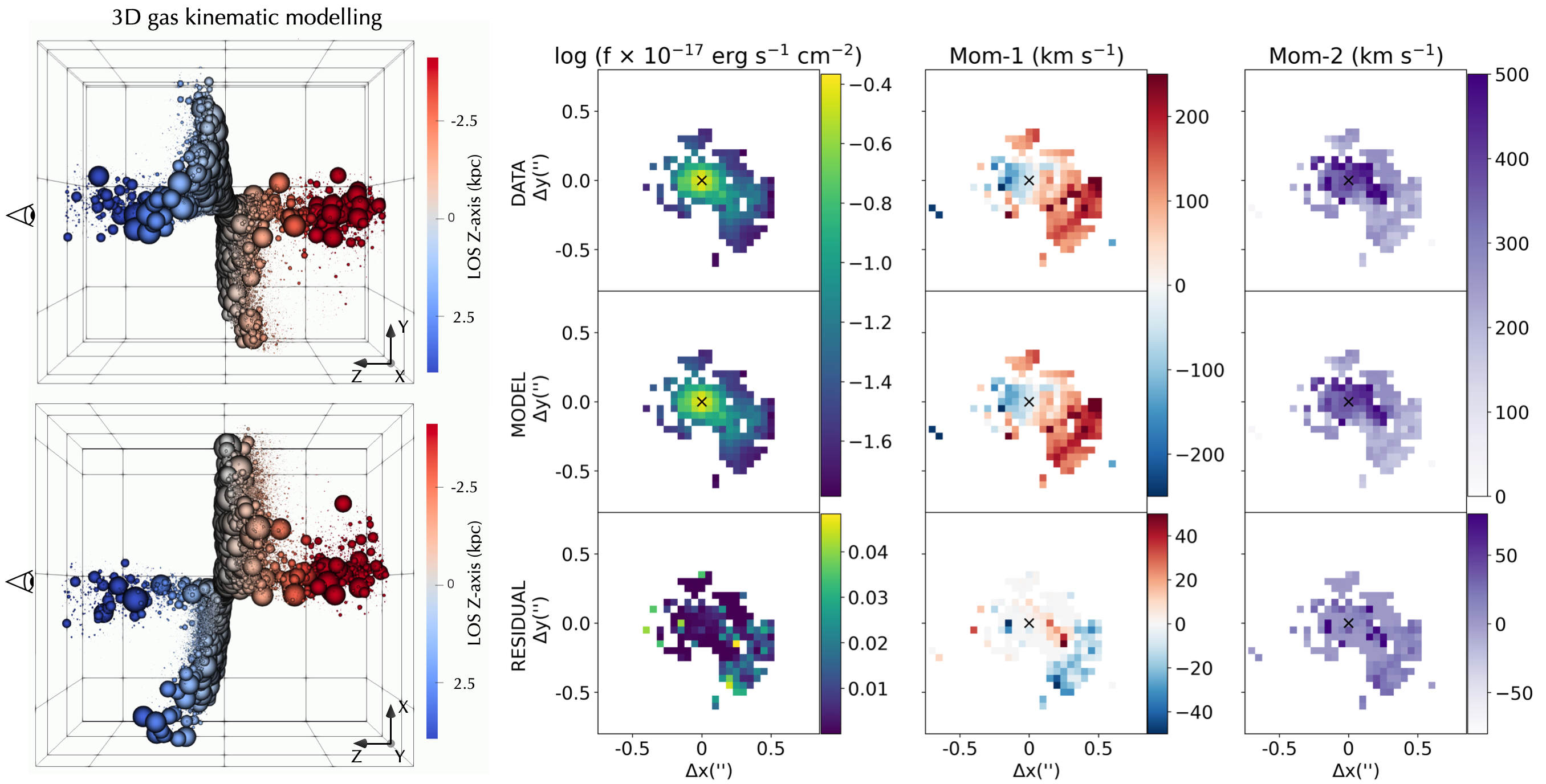}}
\caption{\MOKA  model (left) and moment maps from NIRSpec data and \MOKA  model (right panels). In the 3D representations of the GS133 outflow structure, the XY represents the plane of the sky, while Z axis is the LOS. The observer is positioned on the left, and sees as blueshifted (redshifted) all the approaching (receding) gas, according to the colorbar; largest bubbles identify brightest clumps. The three-by-three panels show the comparison between the flux, Mom-1 and Mom-2 maps of \oiii (top panels) and \MOKA model (middle panels) for the GS133 bi-conical outflow. The bottom panels present the residuals obtained subtracting the model from the data.}\label{fig:moka}
\end{figure*}

\subsection{Energetics of emitting gas outflow}

We computed the mass outflow rate of the ionised gas as inferred from the blueshifted outflow component of \hb, assuming a simplified biconical outflow distributed out to a radius $R_{\rm out}$, following \citet{Cresci2015a}:

\begin{equation}
    \dot M_{\rm{out}}(Em,~{\rm H\beta}) = 8.6 \times \frac{L_{41}({\rm H\beta}) \ v_{\rm{out}}}{n_e\ R_{\rm{out}}} M_{\odot}~ \rm yr^{-1}
\end{equation}

where L$_{41}$(\hb) is the \hb luminosity associated with the outflow component in units of $10^{41}$ erg s$^{-1}$ (corrected for dust extinction), 
$n_e$ is the electron density in cm$^{-3}$, and 
$v_{\rm{out}}$ is the outflow velocity in \kms.  

For comparison, we also computed the mass outflow rates 
employing the \oiii luminosity and assuming the same geometrical configuration:

\begin{equation}
    \dot M_{\rm{out}}(Em,~{\rm [OIII]}) = 0.5 \times \frac{L_{41}({\rm [OIII]}) \ v_{\rm{out}}}{n_e\ R_{\rm{out}} 10^{[O/H]}} M_{\odot}~ \rm yr^{-1},
\end{equation}

where L$_{41}$(\oiii) is the \oiii luminosity associated with the outflow component in units of $10^{41}$ erg s$^{-1}$ (corrected for dust extinction), and 10$^{[O/H]}$ is the metallicity of the outflowing gas in solar units (see also \citealt{CanoDiaz2012}). The kinetic power of the ejected emitting gas can be derived as $\dot E = 1/2 \dot M v^2$: 

\begin{equation}
    \dot E_{\rm{out}}(Em) = 3 \times 10^{35} \left ( \frac{{\dot M}(Em)}{M_\odot~yr^{-1}}\right ) \left ( \frac{v_{\rm out}}{km~s^{-1}} \right ) ^2~erg~ s^{-1}.
\end{equation}

For simplicity, the outflow energetics of the emitting gas were derived assuming a simple single-radius wind, because of the unknown detailed geometrical configuration of the outflow (i.e. opening angles and inclinations of the conical structures, deprojected sizes and velocities). 
The fact that the gas is ionised by the AGN did not allow us to derive metallicity measurements for the outflowing gas, but the high \nii/\ha line ratio ($>0.3$) and stellar mass of the system ($\sim 10^{10}$~M$_\odot$) allows us to reasonably assume a solar metallicity to derive \oiii-based outflow energetics. We also assumed an outflow extension of 3~kpc, and an outflow velocity $v_{\rm out} = 1000$~\kms for both \oiii and \hb, as order of magnitude estimates (based on the observed  distribution of high-$v$ \oiii gas in Fig. \ref{fig:spectroastrometry}, and our multi-Gaussian fit decomposition results reported in Table \ref{tab:OPTproperties}). Similarly, we assumed an electron density $n_e = 1000$~cm$^{-3}$, considering that we could not obtain a measurement for the ejected gas because of the faintness and complexity of the \sii profiles, and the extremely high density derived from the narrow \sii components ($3300\pm 500$~cm$^{-3}$) in the integrated spectrum shown in Fig. \ref{fig:opticalintegratedspectrum}. With these assumptions, we obtained $\dot {\rm M}_{out}$(\oiii)~$= 60$~\Msunyr and $\dot {\rm M}_{out}$(\hb)~$= 200$~\Msunyr, respectively. We note that the \oiii-based estimates are a few times lower than those obtained with \hb, in agreement with previous results suggesting that \oiii provides a lower limit on the ionised gas mass (see e.g. \citealt{Carniani2015,Perna2015a, Marshall2023, Venturi2023teacup}). Therefore, in the following, we refer only to the \hb-based outflow energetics for the emitting gas component of the GS133 outflow.
The inferred outflow mass rate, kinetic power, and momentum power (defined as $\dot P = \dot M v$), are reported in Table \ref{tab:outflowproperties}.
In Sect. \ref{sec:physmech} we compare these quantities with the AGN bolometric luminosity and some theoretical predictions to infer the driving mechanisms of the mini-BAL and \oiii outflows detected in GS133.

\subsection{3D modelling of the outflowing emitting gas}\label{sec:geometry}

Our \textsc{Cloudy} models suggest that the mini-BAL absorbing material, traced by \cii, \civ, and other atomic species detected in the VIMOS spectrum of GS133, extends to kpc scales. Since we observe the mini-BAL directly along the LOS, we aim to determine whether this absorbing outflow is partially mixed with the emitting \oiii gas. 

Despite the fact that the \oiii outflow in GS133 extends over kpc-scales, it only covers a few tens of spaxels in the NIRSpec cube (Fig. \ref{fig:outflowmap}). 
A detailed reconstruction of the 3D geometry of the outflow, as done for targets at lower redshifts (e.g. \citealt{Meena2021, Cresci2023, Ulivi2024b}), would require finer spatial sampling and higher angular resolution than what NIRSpec IFS provides.  
Nonetheless, we used our framework ``Modelling Outflows and Kinematics of AGN in 3D'' (\MOKA; \citealt{Marconcini2023}) to test whether our NIRSpec data are compatible with the presence of a biconical outflow where the approaching side intersects our LOS.

For simplicity, we adopted constant radial velocity profiles for both the approaching and receding components of the \oiii outflow. We also assumed a bi-conical geometry, with a semi-aperture angle of 45$^\circ$, consistent with other AGN-driven outflows studied at lower redshifts (e.g. \citealt{Fischer2014, MullerSanchez2016, Meena2021, Perna2022}). 
For the approaching cone, we required an inclination angle with respect to the LOS in the range [--45$^\circ$, +45$^\circ$], to ensure the overlap with our LOS; for the receding cone, we required an inclination angle in the range [180$^\circ - 45^\circ$, 180$^\circ+45^\circ$]. 

The \oiii outflow is reproduced with \MOKA with an approaching cone with an inclination angle of 40$^\circ$ with respect to the LOS, a position angle of 50$^\circ$\footnote{The position angles are measured anti-clockwise from the north.}, and an intrinsic deprojected outflow velocity of 900~\kms; the receding cone has an inclination angle of $225^\circ$, a position angle of 230$^\circ$, and an intrinsic deprojected outflow velocity of 800~\kms. The two cones extend out to 5 kpc. Figure \ref{fig:moka} shows the 3D representation of the outflow, together with the flux, velocity, and velocity dispersion maps from data, the 3D model, and the residuals; the figure indicates the \MOKA capabilities in recovering the general properties of \oiii outflowing gas with a bi-conical model. 
In particular, a hollow conical best-fit model is obtained, consistent with many other AGN-driven outflows reported in the literature (e.g. \citealt{Marconcini2023} and references therein).
We also note that the \MOKA model necessitates the presence of emission on kiloparsec scales located close to the plane of the sky. This component exhibits relatively low projected velocities and can explain the slightly redshifted emission observed toward the NE and the blueshifted emission toward the SW in the narrow component maps shown in Fig. \ref{fig:diskmap} which cannot be attributed to a rotating pattern.

From this simple \MOKA model, we can derive two main results. On the one hand, the \oiii gas appears to share the same velocity as the low-velocity component in \cii, \civ, and other UV lines ($|v10_{\rm out}| \sim v_{\rm MOKA} \approx 900$~\kms); this suggests that both the absorbing gas and part of the emitting \oiii could lie along our LOS and may be associated with the same outflow, possibly even physically mixed.
On the other hand, the second kinematic component detected in absorption in \civ and other HILs moves at $v_{\rm out}\sim -1900$~\kms (see Fig. \ref{fig:Voigtfit}); this component is not detected in the ionised emitting gas, and could therefore have slightly different physical (e.g. composition, ionization state) or structural (e.g. spatial location) properties. 
Alternatively, this very high-velocity component may be too faint to be detected in emission.

\subsection{Physical mechanisms for the wind propagation}\label{sec:physmech}

We compared our inferred values of the total outflow kinetic power with the AGN bolometric luminosity and the expected kinetic power ascribed to stellar processes. The AGN bolometric luminosity can be constrained from the X-ray emission, applying a \citet{Duras2020} bolometric correction of 260 to the absorption corrected 2 -- 10~keV luminosity from \citet{Li2019}, L$_{\rm bol}= 1.6\times 10^{45}$~\ergs. This value is broadly consistent with the SED-based AGN luminosity inferred by \citet[][L$_{\rm bol}= 4.9\times 10^{45}$~\ergs]{Guo2021}, and by Circosta et al., in prep. ($6.7\times 10^{44}$~\ergs). However, these values are significantly lower than those inferred from the outflow component of the optical emission lines: from \hb flux, applying the \citet{Netzer2019} bolometric correction, L$_{\rm bol} ({\rm H\beta})= 4.5\times 10^{46}$~\ergs, and from \oiii flux, applying the \citet{Heckman2004} correction, L$_{\rm bol}$(\oiii) $= 3\times 10^{46}$~\ergs (even higher values would be obtained applying a dust extinction correction, and considering the total line profiles, which however could be contaminated by star-formation processes). 
Since the X-ray-based value is in between those from the SED analysis, in the following, we refer to the former as our fiducial estimate.  

The kinetic power associated with the optical outflow is $3\times 10^{42}$~\ergs, hence $\sim 0.2$\% of the radiative luminosity of the AGN, consistent with the energetics of other AGN in the literature (e.g. \citealt{Harrison2018}). 
Comparable values can be obtained  for the mini-BAL considering a kpc-scale extension: $\dot E_{\rm out}/L_{\rm bol} \sim 0.1$\% for R$_{\rm out}=1$~kpc, and $\sim 1\%$ for R$_{\rm out}=10$~kpc.  

Following \citet{Veilleux2005}, we assumed an expected kinetic power ascribed to stellar processes proportional to the SFR ($20$~\Msunyr, from \citealt{Guo2020}, $<70$~\Msunyr, from Circosta et al., in prep.), and concluded that the outflow kinetic power of the emitting gas cannot be associated with star formation driven winds, as they would require an extreme coupling between the stellar processes and the observed winds, of the order of $\approx 100$\% or more.

The momentum power inferred for the optical emitting gas is $\dot P_{\rm H\beta} = 1\times 10^{36}$~dyne, with a momentum rate $\dot P_{\rm H\beta}$/(L$_{\rm bol}$/c)~$= 24$. For the UV absorbing gas, we obtained $(2-20)\times 10^{35}$~dyne, resulting in $\dot P_{\rm UV}$/(L$_{\rm bol}$/c)~$=4-40$, with lowest values occurring for a mini-BAL with R$_{\rm out}$ = 1~kpc and highest for R$_{\rm out}$ = 10~kpc (still compatible with measurements of other AGN in the literature, see e.g. Fig. 3 in \citealt{Bischetti2024}).

A momentum flux rate obtained under the assumption of a very compact mini-BAL is far smaller than the moment flux rate of the \hb outflow ($\dot P_{\rm UV}$/(L$_{\rm bol}$/c)~$= 0.004$ for R$_{\rm out} = 1$~pc), and hence than what is expected for a momentum conserving wind (e.g. \citealt{Fiore2017}). If the AGN outflow is energy conserving, it is possible that the mini-BAL is radiatively launched with small momentum flux in the nucleus, and then boosted to $\dot P_{\rm H\beta}$/(L$_{\rm bol}$/c)~$\sim 24$ at larger distances (e.g. \citealt{FaucherGiguere2012}). However, such a dramatic increase in momentum flux, by four orders of magnitude, is highly uncommon (see Fig. 8 in \citealt{Tozzi2021}). 
This further supports our \textsc{Cloudy} model results indicating that the mini-BAL is located at kpc scales.

Summarising, the computation of outflow energetics allowed us to obtain an independent (but still indirect) confirmation of the kpc-scale location of the mini-BAL; the modest momentum bursts inferred for optical and UV transitions, of the order of 2 to 40, are more likely compatible with energy conserving winds rather than momentum driven ones (e.g. \citealt{King2015}), also taking into account that with current data we cannot probe the neutral and molecular outflow phases which could significantly contribute to the outflow energetics on kpc-scales (e.g. \citealt{Brusa2018, Cresci2023, Davies2024, DEugenio2023}).

\begin{table}[h]
\tabcolsep 20.5pt 
\centering
\caption{Mini-BAL and \oiii outflow energetics.}%
\begin{tabular}{lc}
\hline

Measurement  & Value \\
\hline
\hline

L$_{\rm bol}$(X-ray) & $1.6\times 10^{45}$~\ergs \\
L$_{\rm bol}$(\hb) & $4.5\times 10^{46}$~\ergs \\
L$_{\rm bol}$(\oiii) & $3.0\times 10^{46}$~\ergs \\
L$_{\rm bol}$(SED)$^\dagger$ & $6.7\times 10^{44}$~\ergs \\
L$_{\rm bol}$(SED)$^\clubsuit$ & $4.9\times 10^{45}$~\ergs \\
\hline
$\dot M$(\hb) & $200$~\Msunyr \\
$\dot E$(\hb) & $3\times 10^{43}$~\ergs \\
$\dot E$(\hb)/L$_{\rm bol}$ & 2\% \\
$\dot P$(\hb) & $10^{36}$~dyne \\
$\dot P$(\hb)/(L$_{\rm bol}$/c) & 24 \\

\hline

\hline
$\dot M$(UV)$_{1~{\rm kpc}}$ & $20$~\Msunyr \\
$\dot E$(UV)$_{1~{\rm kpc}}$ & $2\times 10^{43}$~\ergs \\
$\dot E$(UV)$_{1~{\rm kpc}}$/L$_{\rm bol}$ & 0.1\% \\
$\dot P$(UV)$_{1~{\rm kpc}}$ & $2\times 10^{35}$~dyne \\
$\dot P$(UV)$_{1~{\rm kpc}}$/(L$_{\rm bol}$/c)& 4\\

\hline
$\dot M$(UV)$_{10~{\rm kpc}}$ & $200$~\Msunyr \\
$\dot E$(UV)$_{10~{\rm kpc}}$ & $2\times 10^{44}$~\ergs \\
$\dot E$(UV)$_{10~{\rm kpc}}$/L$_{\rm bol}$ & 1\% \\
$\dot P$(UV)$_{10~{\rm kpc}}$ & $2\times 10^{36}$~dyne \\
$\dot P$(UV)$_{10~{\rm kpc}}$/(L$_{\rm bol}$/c)& 40\\

\hline

\hline

\hline

\end{tabular} 
\tablefoot{Kinetic power coupling efficiencies and momentum bursts are computed with the bolometric luminosity inferred from X-rays, with a bolometric correction of 260 (see Sect. \ref{sec:energetics}); for the mini-BAL, we considered two potential outflow extents, 1 and 10 kpc, which align with the range determined from the \textsc{Cloudy} models in Sect.~\ref{sec:cloudymodels}. ($^\dagger$) SED fitting value from Circosta et al., in prep. ($^\clubsuit$) SED fitting value from \cite{Guo2021}.}   
\label{tab:outflowproperties}
\end{table}

\begin{figure}[!t]
\centering
\includegraphics[width=0.47\textwidth]{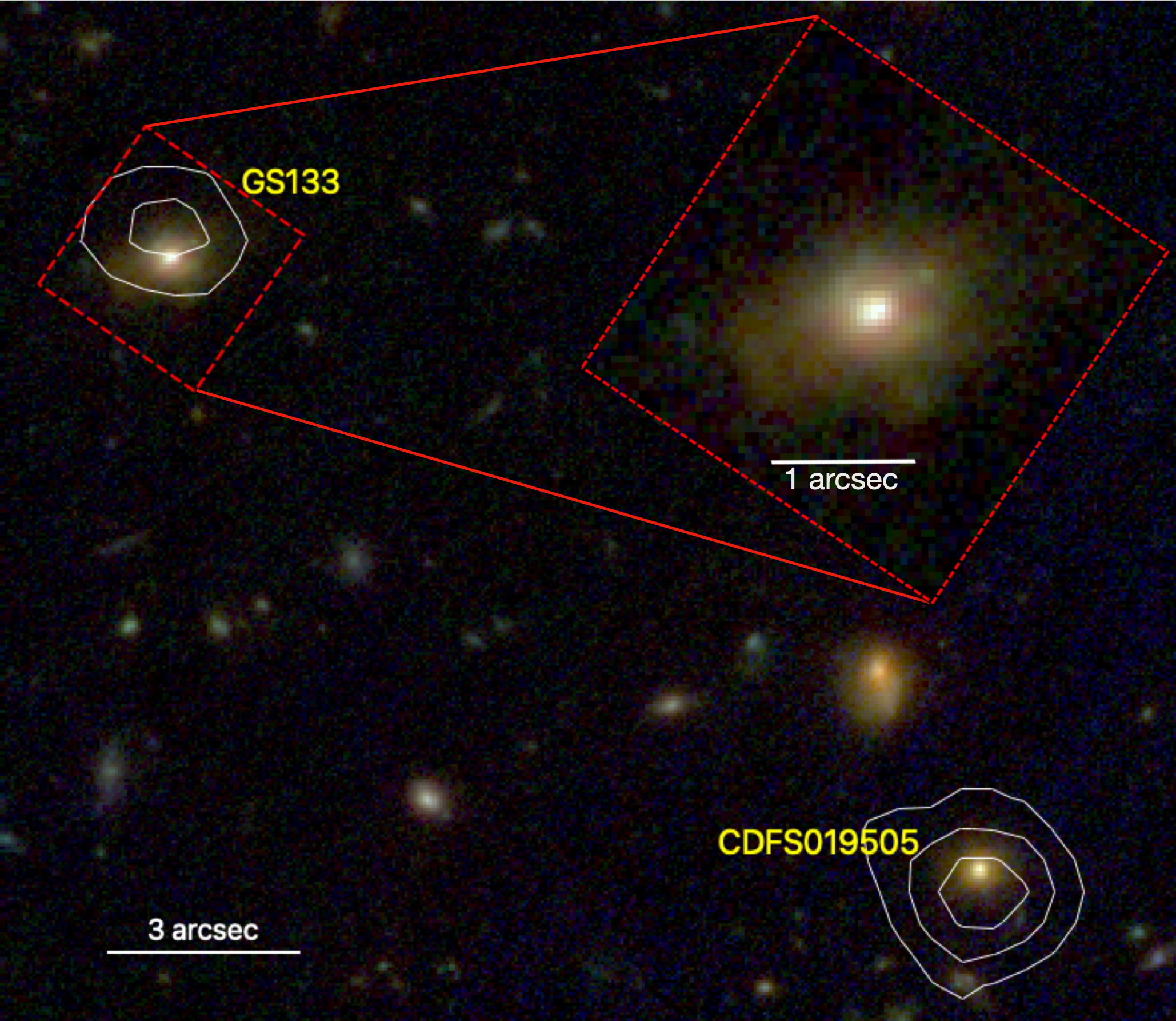}
\caption{ JWST/NIRCam F115W/F200W/F444W cutout showing the close environment of GS133. The red dashed box marks the NIRSpec IFS FOV, while the white contours show the Chandra 0.5--7~keV emission from \cite{Luo2017}, highlighting the position of two X-ray AGN with a projected separation of $\sim 130$~kpc and a velocity offset of $\sim 5000$~\kms ($\Delta z = 0.08$). The apparent offset between the NIR and X-ray positions can be attributed to Chandra's low angular resolution and the fact that the Chandra image is composed of multiple observations taken with varying aim points and roll angles. The inset shows a zoom-in on GS133. The images are oriented north up, east left. }\label{fig:cutout}
\end{figure}

\section{Environment}\label{sec:environment}

NIRSpec IFS data do not reveal any close companions or line emitters around GS133, in contrast with several $z>3$ AGN and QSOs recently observed with JWST, many of which show multiple nearby companions (e.g., \citealt{Wylezalek2022, Marshall2023, Marshall2024, Perna2023,Perna2023b, Ubler2023, Ubler2024a, Ubler2024b, Kashino2023,Yang2023}). Available JWST/NIRCam images (program ID 3990, PI: T. Morishita) show some extended emission towards E-SE associated with the GS133 host, possibly indicating merger features or a disturbed morphology, and faint compact sources outside of the NIRSpec FOV. Figure \ref{fig:cutout} shows the composite F115W/F200W/F444W cutout, highlighting the GS133 extended features. However, these faint features remain undetected in the  spectroscopic data, which are too shallow ($\lesssim 1$~hr-integration) in comparison with NIRCam imaging (e.g. 6 hr-integration in the red band F444W).

Interestingly, GS133 is located near another AGN, identified as CDFS019505 (in the bottom right of Fig. \ref{fig:cutout}), targeted by VLT/VIMOS as part of the VANDELS project and classified as a narrow-line AGN in \citet{Mascia2023}. CDFS019505 was observed with a total integration time of 40 hours, and its spectroscopic redshift, $z = 3.396$, is considered insecure, as it is based on a single emission line at $\sim 5350\AA$, assumed to be Ly$\alpha$ (see Fig. 7, bottom panel, in \citealt{Carnall2020}). On the other hand, this redshift is close to the photometric estimates (e.g. $z_{phot} = 3.417$ in \citealt{Hsu2014}, $z_{phot} = 3.49_{-0.27}^{+0.09}$ in \citealt{Carnall2020}), and it is supported by the detection of continuum emission at observed wavelengths $> 5350\AA$. SED analysis by \citet{Carnall2020} classified CDFS019505 as a quiescent galaxy, requiring an escape fraction $f_{esc, \ {\rm Ly\alpha}} >0.1$, with a Ly$\alpha$-based SFR of $0.28\pm 0.05/f_{esc, \ {\rm Ly\alpha}}$~\Msunyr (similar SED-based value is reported in \citealt{Guo2020}, SFR = 1 \Msunyr).
This source is also classified as an AGN on the basis of its X-ray luminosity, with 140 full (0.5-7~keV) aperture-corrected source counts, an intrinsic column density N$_{\rm H} = 10^{24}$~cm$^{-2}$, and an absorption-corrected intrinsic luminosity L$_{2-10~keV} =  10^{44}$~\ergs (\citealt{Li2019}). 

With a projected separation of 17\arcsec\ (i.e. $\sim 130$~kpc) and a velocity offset of $\sim -5000$~\kms relative to GS133, CDFS019505 displays a significantly larger separation and redshift difference\footnote{We assumed the spectroscopic redshift $z=3.396$ for CDFS019505.} than is typical for close AGN pairs, or ``dual AGN'' (e.g., \citealt{DeRosa2019, Mannucci2023, Scialpi2024, Chen2024arXiv240916351C}). 
\cite{Perna2023b} reported the discovery of three dual AGN systems in the GA-NIFS sample, with much closer projected separations of 5--10~kpc and velocity offsets of a few hundred \kms, suggesting close interactions likely to result in mergers. 
In contrast, the spatial and redshift separation between GS133 and CDFS019505 makes a direct interaction between these two AGN improbable at the time of observation. Nonetheless, the  identification of another AGN pair among the 12 AGN systems at $z\sim 3$ targeted in GA-NIFS further supports the hypothesis that AGN at high redshift tend to reside in dense environments (e.g. \citealt{Karouzos2014}), potentially hosting multiple accreting SMBHs within close proximity (see also e.g. \citealt{Coogan2018, Pensabene2024}, Zamora et al., in prep.).

\section{Conclusions}\label{sec:conclusions}

We have presented JWST NIRSpec IFS data of the Compton thick quasar GS133 at $z = 3.46$. These observations allowed us to map the extension of the quasar host 
traced by rest-frame optical emission lines. NIRSpec data have been complemented with 1D spectroscopic data from VLT/VIMOS, covering the rest-frame UV lines, and NIRCam images, to explore the GS133 environment.
These are our main results.

We identified mini-BAL features in multiple transitions covered by VIMOS data, including \nv, \siiv, \civ, \cii. The column densities derived from VoigtFit for individual metal species span a narrow range of values, log (N/cm$^{-2}$)~=~18.3--19.4. Higher ionisation lines (e.g. \civ and \siiv) are broader than the lower ionisation lines (\cii), and show two kinematic components, shifted by $-800$~\kms and $-1900$~\kms respectively.  
The lower-ionization lines only reveal the $-800$~\kms component.  Both kinematic components are tentatively detected (S/N~$\sim 2$) in absorption in the \ha line observed with NIRSpec.

Using single-cloud \textsc{Cloudy} models, we derived hydrogen density and column density estimates for the mini-BALs to constrain the distance of the absorbing material from the AGN, and, therefore, the outflow energetics: log (N$_{\rm H}$/cm$^{-2}$) = 19.4-20.3, log (n$_{\rm H}$/cm$^{-3}$) = 1-3, with $R_{out} =1-10$~kpc. Outflow energetics were derived assuming a thin-shell approximation, and accounting for uncertainties in the best-fit \textsc{Cloudy} parameters: we obtained a outflow mass rate of 20-200 \Msunyr, and a kinetic power in the range of $2-20\times 10^{43}$~\ergs.

Spaxel-by-spaxel analysis of NIRSpec data allowed us to perform emission lines kinematic decompositions. We revealed two distinct components in the host galaxy of GS133: the first component likely traces a rotating disk with a dynamical mass $\sim 2\times 10^{10}$~M$_\odot$. The second component corresponds to a galaxy-wide, bi-conical outflow, with a velocity of $\sim \pm 1000$~\kms and a projected radius of $\sim 3$~kpc.

The inferred outflow velocity for the \oiii line ($W80_{\rm out}\sim 1300$~\kms, $v10_{\rm out} \sim -900$~\kms) aligns with the general L$_{\rm AGN}-v_{\rm out}$ trends reported in the literature (\citealt{ Musiimenta2023}), placing GS133 at the higher end of the velocity distribution. This is consistent with findings by  \citet{Tozzi2024} showing that CT AGN tend to exhibit faster outflows than unobscured AGN at fixed L$_{\rm bol}$.

We derived emitting-gas outflow energetics assuming a simple bi-conical configuration: the outflow mass rate is 200~\Msunyr, and the kinetic power is $3\times 10^{43}$~\ergs, comparable to those inferred for the mini-BALs in the UV regime. 
The energetics of both absorbing and emitting gas  suggest that the outflow is driven by energy conserving winds, with $\dot E$/L$_{\rm bol}$ = 0.1--2\%, $\dot P$/(L$_{\rm bol}$/c) = 4--40, and mass loading factor $\dot M/$SFR$~= 1-10$.

We explored the potential overlap of absorbing and emitting outflows using \MOKA, modelling the \oiii outflow with a bi-conical structure intersecting our LOS. The \MOKA model fits the data well with the following best-fit parameters: a semi-aperture angle of 45$^\circ$, a radius of 5~kpc, a constant radial velocity of 800--900~\kms, and an inclination of 40$^\circ$ with respect to our LOS, hence with an approaching cone overlapping the LOS.  
Similarities in velocity, location of the absorbing gas, and LOS extension of the emitting gas, suggest at least partial mixing between the mini-BAL and \oiii outflows. However, the faster mini-BAL component seen in \civ and other high-ionisation lines is not present in \oiii, suggesting that this UV component might have different properties, for instance, in terms of gas composition and spatial location.

NIRSpec IFS data reveal no close companions or line emitters around GS133 within the $\sim 3\arcsec \times 3\arcsec$\ FOV. However, JWST/NIRCam images suggest possible merger features that could help explain the Compton thick nature of this AGN and the onset of outflow episodes. 
We also reported the presence of an X-ray-detected AGN at $\sim 130$~kpc and a velocity offset of $\sim -5000$~\kms from GS133. These results are aligned with recent JWST findings that high-$z$ AGN commonly reside in dense environments.

This comprehensive study demonstrates the power of combining UV and optical data to probe the complex outflows in high-redshift AGN, providing insights into their origin and associated feedback processes. NIRSpec data allowed us to spatially resolve both unperturbed and ejected emitting gas, and obtain crucial information about the 3D configuration of a bi-conical outflow at $z\sim 3.5$. With less than 1 hour of total exposure, JWST NIRSpec IFS provided essential complementary data to the extremely deep VLT/VIMOS observations (41-hour integration), demonstrating JWST’s ability to efficiently capture detailed outflow properties.

\section*{Acknowledgements}


MP, SA, BRP, and IL acknowledge support from the research project PID2021-127718NB-I00 of the Spanish Ministry of Science and Innovation/State Agency of Research (MCIN/AEI/10.13039/501100011033). 
%
IL acknowledges support from grant PRIN-MUR 2020ACSP5K\_002 financed by European Union – Next Generation EU.
RM acknowledges support by the Science and Technology Facilities Council (STFC), by the ERC Advanced Grant 695671 ``QUENCH'', and by the UKRI Frontier Research grant RISEandFALL; RM is further supported by a research professorship from the Royal Society.
AJB acknowledges funding from the ``First Galaxies'' Advanced Grant from the European Research Council (ERC) under the European Union’s Horizon 2020 research and innovation program (Grant agreement No. 789056).
MP, GC and EB acknowledge the support of the INAF Large Grant 2022 "The metal circle: a new sharp view of the baryon cycle up to Cosmic Dawn with the latest generation IFU facilities". 
H\"U acknowledges support through the ERC Starting Grant 101164796 ``APEX''.

%


\bibliographystyle{aa}
\bibliography{gs133.bib}

\begin{appendix}
\section{Spaxel-by-spaxel multi-Gaussian fit of the NIRSpec datacube}

\begin{figure*}[!htb]
\centering
\includegraphics[width=0.8\textwidth]{{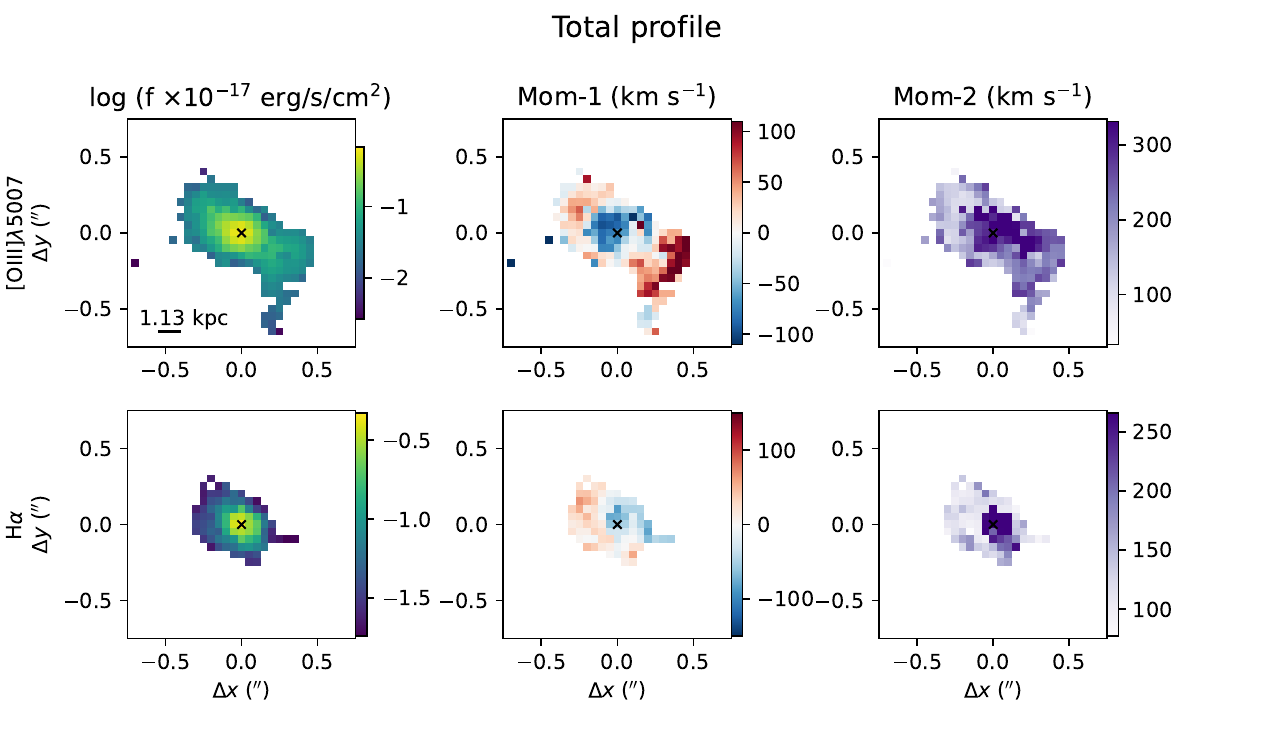}}

\caption{Total profiles \oiii (top) and \ha (bottom) flux, moment-1 and moment-2 maps.
A S/N cut of 4 has been applied to generate the maps. The disk and biconical outflow components identified in Figs. \ref{fig:diskmap} and \ref{fig:outflowmap}, respectively, are here less evident. \ha is less affected by the outflow, and its moment-1 still reveal a NW-SE gradient likely associated with a disk rotation. }\label{fig:totalprofilemaps}
\end{figure*}

\section{Spatially integrated spectra of off-nuclear approaching and receding gas}

\begin{table}[h]
\centering
\caption{Optical emission line properties from the NIRSpec integrated spectrum of NW cone (blueshifted outflow).}%
\begin{tabular}{lc}
\hline

Measurement  & Value\\
\hline
\hline


$\Delta v_{sys}$ & $49\pm 3$\\
$\sigma_{sys}$ & $97\pm 7$\\

\hline


$\Delta v_{out}$ & $-75_{-26}^{+32}$~\kms\\
$\sigma_{out}$ & $265_{-18}^{+28}$~\kms \\

$W80_{\rm out}$ & $688_{-30}^{+58}$~\kms\\
$v10_{\rm out}$ & $-440\pm 60$~\kms\\
$v90_{\rm out}$ & $250\pm 10$~\kms\\

\hline

$\Delta v_{tot}$ & $-2\pm 4$~\kms\\
$\sigma_{tot}$ & $196\pm10$~\kms \\

$W80_{\rm tot}$ & $480\pm 15$~\kms\\
$v10_{\rm tot}$ & $-290\pm 10$~\kms\\
$v90_{\rm tot}$ & $190\pm 10$~\kms\\

\hline

\end{tabular} 
\tablefoot{All velocity measurements refer to the \oiii line.
}
\label{tab:OPTproperties2}
\end{table}

\begin{table}[h]
\centering
\caption{Optical emission line properties from the NIRSpec integrated spectrum of SW cone (redshifted outflow).}%
\begin{tabular}{lc}
\hline

Measurement  & Value\\
\hline
\hline


$\Delta v_{sys}$ & $-14\pm 2$\\
$\sigma_{sys}$ & $137\pm 3$\\

\hline


$\Delta v_{out}$ & $96_{-3}^{+13}$~\kms\\
$\sigma_{out}$ & $311_{-3}^{+8}$~\kms \\

$W80_{\rm out}$ & $840_{-10}^{+18}$~\kms\\
$v10_{\rm out}$ & $-310_{-3}^{+30}$~\kms\\
$v90_{\rm out}$ & $485\pm 10$~\kms\\

\hline

$\Delta v_{tot}$ & $53_{-2}^{+6}$~\kms\\
$\sigma_{tot}$ & $260\pm5$~\kms \\

$W80_{\rm tot}$ & $660\pm 20$~\kms\\
$v10_{\rm tot}$ & $-290\pm 10$~\kms\\
$v90_{\rm tot}$ & $360\pm 10$~\kms\\

\hline

\end{tabular} 
\tablefoot{All velocity measurements refer to the \oiii line.}
\label{tab:OPTproperties3}
\end{table}

\begin{figure*}[!htb]
\centering
\includegraphics[width=\textwidth]{{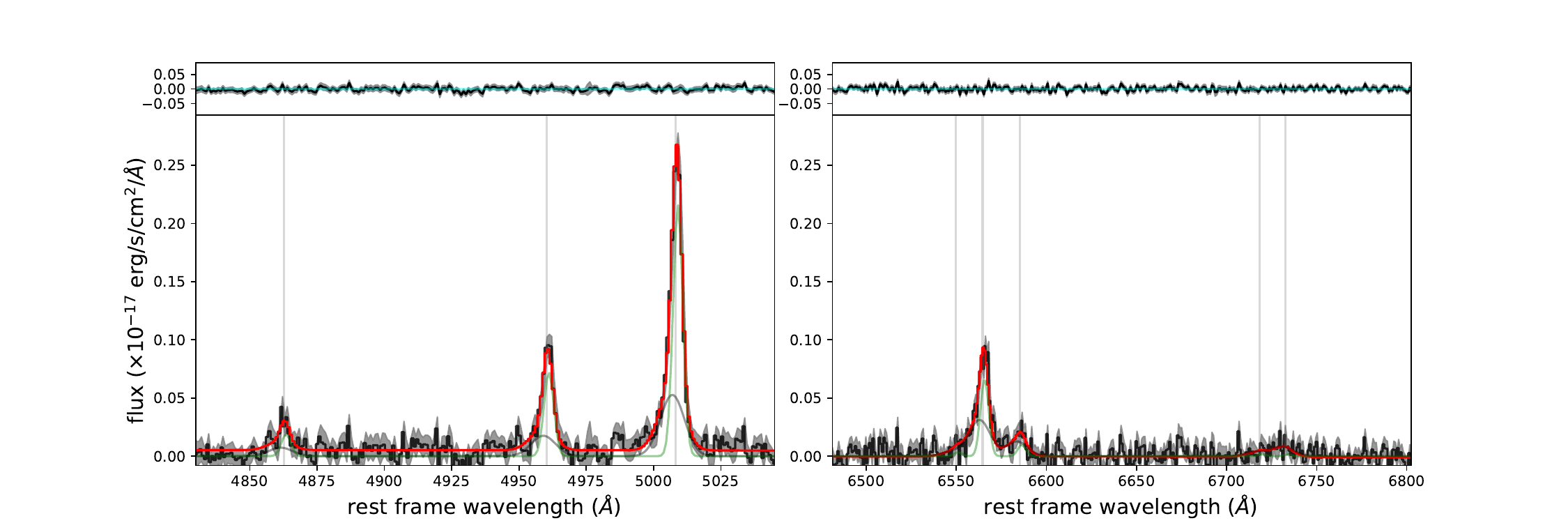}}
\caption{ NIRSpec integrated spectrum of NE cone, showing the blueshifted outflow. The black curve identifies the integrated spectrum, integrated over a circular region identified by the light-blue crescent moon in the inset in Fig. \ref{fig:BPT}; the total, multi-component best fit curve is in red, while the green and and grey Gaussian components show the systemic and outflow, respectively. The fit residuals are reported in the top panels. The most prominent emission lines are marked with grey vertical lines. }\label{fig:NEcone}
\end{figure*}

\begin{figure*}[!htb]
\centering
\includegraphics[width=\textwidth]{{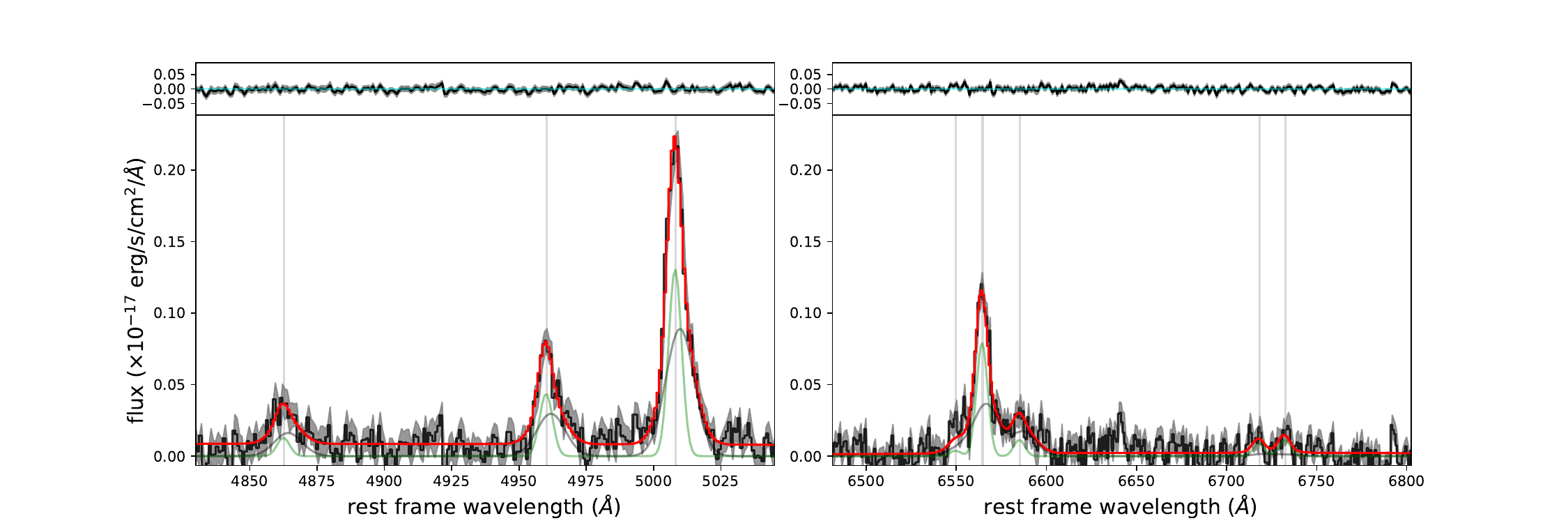}}
\caption{ NIRSpec integrated spectrum of SW cone, showing the redshifted outflow. The black curve identifies the integrated spectrum, integrated over a circular region identified by the light-red crescent moon in the inset in Fig. \ref{fig:BPT}; the total, multi-component best fit curve is in red, while the green and and grey Gaussian components show the systemic and outflow, respectively. The fit residuals are reported in the top panels. The most prominent emission lines are marked with grey vertical lines. }\label{fig:SWcone}
\end{figure*}

\end{appendix}

\end{document}